%% file: draft.tex
\documentclass[useAMS,usenatbib]{mnras}
\usepackage{mn2e-breakabs}
\usepackage{epsfig}
\usepackage[usenames,dvipsnames]{xcolor}
\usepackage{url}
\usepackage{subfigure}
\usepackage[english]{babel}
\usepackage{times,graphicx,amsmath,amsfonts,amssymb,aas_macros,epstopdf,hyperref}
\usepackage[normalem]{ulem}
\usepackage[T1]{fontenc}
\usepackage{multirow}
\usepackage{ae,aecompl}
\usepackage{xspace} 
\usepackage{units}
\usepackage{wasysym}
\usepackage{newtxtext,newtxmath}
\usepackage{arydshln}
\usepackage{adjustbox}
\usepackage{arydshln}

\definecolor{ultramarine}{rgb}{0.07, 0.04, 0.56}
\definecolor{cadmiumgreen}{rgb}{0.0, 0.42, 0.24}
\definecolor{indigo}{rgb}{0.0, 0.25, 0.42}

\hypersetup{
    colorlinks=true,
    linkcolor=blue,
    citecolor=blue,
    urlcolor=blue,
}

\newcommand{\tc}{\color{black}}

\newcommand{\remove}[1]{}

\def\etal{{\frenchspacing\it et al.}}
\def\ie{{\frenchspacing\it i.e.}}
\def\eg{{\frenchspacing\it e.g.}}
\def\etc{{\frenchspacing\it etc.}}

\def\be{\begin{equation}}
\def\ee{\end{equation}}
\def\ba{\begin{eqnarray}}
\def\ea{\end{eqnarray}}

\def\nn{\nonumber}
\def\dd{{\updelta\updelta}}
\def\vv{{\uptheta\uptheta}}
\def\dt{{\updelta\uptheta}}

\def\PLL{P_{\rm{LL}}}
\def\PEE{P_{\rm{EE}}}
\def\PLE{P_{\rm{LE}}}
\def\wiA{{w}^{\rm A}_i}
\def\wiB{{w}^{\rm B}_i}

\def\d{{\rm d}}

\def\rd{r_{\rm d}}

\def\zp{z_{\rm p}}
\def\zL{z_{\rm L}}
\def\zE{z_{\rm E}}
\def\zX{z_{\rm X}}

\def\PlX{P_{\ell}^{\rm X}}

\def\r{{\bf r}}

\def\kunit{{h \ {\rm Mpc}^{-1}}}
\def\abot{{\alpha_{\bot}}}
\def\apar{{\alpha_{\|}}}
\def\fs{{f\sigma_8}}

\def\bsILN{{b_{\rm 1,N}^{\rm L}\sigma_8}}
\def\bsILS{{b_{\rm 1,S}^{\rm L}\sigma_8}}
\def\bsIEN{{b_{\rm 1,N}^{\rm E}\sigma_8}}
\def\bsIES{{b_{\rm 1,S}^{\rm E}\sigma_8}}

\def\bsILNz{{b_{\rm 1,N}\sigma_8}(z=0.70)}
\def\bsILSz{{b_{\rm 1,S}\sigma_8}(z=0.70)}
\def\bsIENz{{b_{\rm 1,N}\sigma_8}(z=0.845)}
\def\bsIESz{{b_{\rm 1,S}\sigma_8}(z=0.845)}

\def\zeff{{z_{\rm eff}}}

\def\LLLL{{\rm LLLL}}
\def\EEEE{{\rm EEEE}}

\def\LELE{{\rm LELE}}

\def\LLEE{{\rm LLEE}}

\def\LLLE{{\rm LLLE}}

\def\EEEL{{\rm EEEL}}

\def\AAAA{{\rm AAAA}}
\def\ABAB{{\rm ABAB}}
\def\AABB{{\rm AABB}}
\def\AAAB{{\rm AAAB}}

\def\L{{\rm L}}
\def\E{{\rm E}}
\def\Q{{\rm Q}}
\def\A{{\rm A}}
\def\B{{\rm B}}

\def\f{\frac}
\def\l{\left}
\def\r{\right}

\def\DM{D_{\rm M}}
\def\DH{D_{\rm H}}

\newcommand{\bfv}{\mbox{\boldmath$v$}}

\newcommand{\bfx}{\mbox{\boldmath$x$}}
\newcommand{\bfk}{\mbox{\boldmath$k$}}
\newcommand{\bfp}{\mbox{\boldmath$p$}}
\newcommand{\bfq}{\mbox{\boldmath$q$}}
\newcommand{\bfr}{\mbox{\boldmath$r$}}
\newcommand{\bfs}{\mbox{\boldmath$s$}}
\newcommand{\bfu}{\mbox{\boldmath$u$}}

\newcommand{\deltas}{\delta^{\rm(S)}}

\newcommand{\deltaA}{\delta_{\rm A}}
\newcommand{\deltaB}{\delta_{\rm B}}

\newcommand{\Pkaiser}{\widetilde{P}_{\rm\scriptscriptstyle Kaiser}}
\newcommand{\Aterm}{\widetilde{A}}
\newcommand{\Bterm}{\widetilde{B}}

\newcommand{\bispecAterm}{\widetilde{B}_{\sigma}}

\newcommand{\sigmav}{\tilde{\sigma}_{\rm v}}

\newcommand{\bA}{b_{\rm A}}
\newcommand{\bB}{b_{\rm B}}
\newcommand{\cA}{c_{\rm A}}
\newcommand{\cB}{c_{\rm B}}

\newcommand{\bIA}{b_1^{\rm A}}
\newcommand{\bIIA}{b_2^{\rm A}}
\newcommand{\bIIIA}{b_{\rm 3nl}^{\rm A}}
\newcommand{\bIB}{b_1^{\rm B}}
\newcommand{\bIIB}{b_2^{\rm B}}
\newcommand{\bIIIB}{b_{\rm 3nl}^{\rm B}}
\newcommand{\bsA}{b_{\rm s2}^{\rm A}}
\newcommand{\bsB}{b_{\rm s2}^{\rm B}}
\newcommand{\bs}{b_{\rm s2}}
\newcommand{\bI}{b_1}
\newcommand{\bII}{b_2}
\newcommand{\bIII}{b_{\rm 3nl}}

\begin{document}

\title[A multi-tracer analysis of the eBOSS DR16 sample]{The Completed SDSS-IV extended Baryon Oscillation Spectroscopic Survey: a multi-tracer analysis in Fourier space for measuring the cosmic structure growth and expansion rate}

\input{authors}

\date{Accepted XXX. Received YYY; in original form ZZZ}
\pubyear{2020}

\label{firstpage}
\pagerange{\pageref{firstpage}--\pageref{lastpage}}

\maketitle

\begin{abstract} 
We perform a joint BAO and RSD analysis using the eBOSS DR16 LRG and ELG samples in the redshift range of $z\in[0.6,1.1]$, and detect a RSD signal from the cross power spectrum at a $\sim4\sigma$ confidence level, \ie, $\fs=0.317\pm0.080$ at $\zeff=0.77$. Based on the chained power spectrum, which is a new development in this work to mitigate the angular systematics, we measure the BAO distances and growth rate simultaneously at two effective redshifts, namely, $\DM/\rd \ (z=0.70)=17.96\pm0.51, \ \DH/\rd \ (z=0.70)=21.22\pm1.20, \ \fs \ (z=0.70) =0.43\pm0.05$, and $\DM/\rd \ (z=0.845)=18.90\pm0.78, \ \DH/\rd \ (z=0.845)=20.91\pm2.86, \ \fs \ (z=0.845) =0.30\pm0.08$. Combined with BAO measurements including those from the eBOSS DR16 QSO and Lyman-$\alpha$ sample, our measurement has raised the significance level of a nonzero $\Omega_{\rm \Lambda}$ to $\sim11\sigma$. The data product of this work is publicly available at \url{https://github.com/icosmology/eBOSS_DR16_LRGxELG} and \url{https://www.sdss.org/science/final-bao-and-rsd-measurements/}  
\end{abstract}


\input{introduction}
\input{data}

\input{method}
\input{result}

\input{conclusion}
\input{thanks}

\bibliographystyle{mnras}
\bibliography{draft}
\bsp
\input{appendix}

\label{lastpage}
\end{document}

%% file: authors.tex
\author[Zhao \etal]{
\parbox{\textwidth}{
Gong-Bo Zhao$^{1,2}$\thanks{Email: \url{gbzhao@nao.cas.cn}}, 
Yuting Wang$^{1}$,
Atsushi Taruya$^{3,4}$,
Weibing Zhang$^{2,1}$,
H\'ector Gil-Mar\'{\i}n$^{5}$,
Arnaud de~Mattia$^{6}$,
Ashley J. Ross$^{7}$,
Anand Raichoor$^{8}$,
Cheng Zhao$^{8}$,
Will J. Percival$^{9}$,
Shadab Alam$^{10}$,
Julian E. Bautista$^{11}$,
Etienne Burtin$^{6}$,
Chia-Hsun Chuang$^{12}$,
Kyle S. Dawson$^{13}$,
Jiamin Hou$^{14}$,
Jean-Paul Kneib$^{8}$,
Kazuya Koyama$^{11}$,
H{\'e}lion du Mas des Bourboux$^{13}$,
Eva-Maria Mueller$^{15}$,
Jeffrey A. Newman$^{16}$,
John A. Peacock$^{10}$,
Graziano Rossi$^{17}$,
Vanina Ruhlmann-Kleider$^{7}$,
Donald P. Schneider$^{18,19}$,
Arman Shafieloo$^{20}$
}
\vspace*{30pt} \\
$^1$ National Astronomy Observatories, Chinese Academy of Sciences, Beijing, 100012, P.R.China\\
$^2$ University of Chinese Academy of Sciences, Beijing 100049,
China \\
$^3$ Center for Gravitational Physics, Yukawa Institute for Theoretical Physics, Kyoto University, Kyoto 606-8502, Japan \\
$^4$ Kavli Institute for the Physics and Mathematics of the Universe (WPI), The University of Tokyo,\\
Institutes for Advanced Study,
The University of Tokyo, 5-1-5 Kashiwanoha, Kashiwa, Chiba 277-8583, Japan \\
$^{5}$ ICC, University of Barcelona, IEEC-UB, Mart\' i i Franqu\` es, 1, E08028 Barcelona, Spain\\
$^{6}$ IRFU,CEA, Universit\'e Paris-Saclay, F-91191 Gif-sur-Yvette, France\\
$^{7}$ Center for Cosmology and Astro-Particle Physics, Ohio State University, Columbus, Ohio, USA\\
$^{8}$ Institute of Physics, Laboratory of Astrophysics, Ecole Polytechnique F\'ed\'erale de Lausanne (EPFL), Observatoire de Sauverny, 1290 Versoix, Switzerland \\ 
$^{9}$ Waterloo Centre for Astrophysics, Department of Physics and Astronomy, University of Waterloo, Waterloo, ON N2L 3G1, Canada\\
$^{10}$ Institute for Astronomy, University of Edinburgh, Royal Observatory, Blackford Hill, Edinburgh, EH9 3HJ, UK\\
$^{11}$ Institute of Cosmology \& Gravitation, University of Portsmouth, Dennis Sciama Building, Portsmouth, PO1 3FX, UK\\
$^{12}$ Kavli Institute for Particle Astrophysics and Cosmology, Stanford University, 452 Lomita Mall, Stanford, CA 94305, USA\\
$^{13}$ Department Physics and Astronomy, University of Utah, 115 S 1400 E, Salt Lake City, UT 84112, USA\\
$^{14}$ Max-Planck-Institut f\"ur Extraterrestrische Physik, Postfach 1312, Giessenbachstrasse 1, 85748 Garching bei M\"unchen, Germany\\
$^{15}$ Sub-department of Astrophysics, Department of Physics, University of Oxford, Denys Wilkinson Building, Keble Road, Oxford OX1 3RH\\
$^{16}$ PITT PACC, Department of Physics and Astronomy, University of Pittsburgh, Pittsburgh, PA 15260, USA \\
$^{17}$ Department of Physics and Astronomy, Sejong University, Seoul 143-747, Korea\\
$^{18}$ Department of Astronomy and Astrophysics, The Pennsylvania State University, University Park, PA 16802, USA\\
$^{19}$ Institute for Gravitation and the Cosmos, The Pennsylvania State University, University Park, PA 16802, USA\\
$^{20}$ Korea Astronomy and Space Science Institute, 776 Daedeokdae-ro, Yuseong-gu, Daejeon 305-348, Republic of Korea\\
}

%% file: introduction.tex
\section{Introduction}

Large spectroscopic galaxy surveys are one of the key probes of both the expansion history and structure growth of the Universe, thus can in principle break the `dark degeneracy' between scenarios of dark energy (DE) (\eg \ \citealt{DEreview}) and modified gravity (MG) (\eg \ \citealt{MGreview}), which are proposed as possible physical origins of the cosmic acceleration \citep{Riess,Perlmutter}. 

Being almost not clustered, dark energy primarily affects the background expansion of the Universe, which can be probed by Baryonic acoustic oscillations (BAO), a special three-dimensional clustering pattern of galaxies, formed in the early Universe due to interactions between photons and baryons. {\tc The BAO feature was first detected by both the Sloan Digital Sky Survey (SDSS) collaboration \citep{Eisenstein05} and the 2 degree Field Galaxy Redshift Survey (2dFGRS) collaboration in 2005 \citep{2dFBAO}}, and has been extensively investigated by a large number of studies since then.

Modified gravity, on the other hand, can dictate both the expansion and the structure formation of the Universe. After the required tuning for the cosmic acceleration, MG leaves imprints at the perturbation level, \ie, it can alter the history of the structure growth on linear and nonlinear scales. On such scales where the peculiar motion of galaxies is relevant, the redshift space distortions (RSD) can be directly mapped by redshift surveys, as reported by its first detection in 2001 by the 2dFGRS collaboration \citep{2dfrsd}.

Measurements of BAO and RSD from redshift surveys and cosmological implications have been extensively performed \citep{sdss2bao,BAO6dF,6dfrsd,wigglezrsd,GAMA,wigglezbao,mgs,desy1,alam,FBRSD16,Zhaotomo16,Wangtomo16,Wangtomo17,dr14lrg,DR14BAO,HGM18,PZ18,Zheng19,Zhao19}, but most of studies focus on the clustering of a single type of galaxies. This is, however, largely due to the fact that most finished galaxy surveys, including 2dFGRS and SDSS III-BOSS, only target at a single tracer in the same cosmic volume.

The statistical error budget of RSD measurements is dominated by the shot noise and the cosmic variance on small and large scales, respectively. While the former can be in principle reduced by increasing the number densities of the observed tracers, the latter is difficult to suppress, due to the fact that the number of large-scale modes is limited by the survey volume. One possible way to tackle the cosmic variance, however, is to combine multiple tracers with different biases covering the same footprint and redshift range \citep{multi-tracer1,multi-tracer2}. The idea is that by contrasting different tracers of the same underlying density field, the uncertainty of statistics of the density field, which is dominated by the cosmic variance on large scales, can be cancelled out if the shot noise of all the concerning tracers is negligible, yielding a measurement of $R_b$, the ratio of {\it effective} biases, $R_b\equiv b_{\rm eff}^{X}/b_{\rm eff}^{Y}$, between tracers $X$ and $Y$ without cosmic variance. The measured bias is effective because it includes the RSD term, namely, $b_{\rm eff} \supseteq b\left(1+\beta\mu^2\right)$, where $b,\beta$ and $\mu$ are the linear bias, the RSD parameter and the cosine of the angel between the line-of-sight and the pair of tracers, respectively. The effective bias also receives a contribution from the primordial non-Gaussianity parametrized by $f_{\rm NL}$, if $f_{\rm NL}\ne0$. By combining measurements of $R_b$ using various $\mu$ modes, parameters of $\beta$, or $f_{\rm NL}$ can be determined to an arbitrary precision in the ideal case, where the shot-noise is negligible.

It is challenging to run a multi-tracer survey, as different tracers may require different methods of target selection, different treatments of observational systematics, and different tracers have to be observed separately, making it expensive to build and perform. {\tc Alternative options include either creating `multi-tracer' samples from a single-tracer survey by splitting the samples using luminosity or colour \citep{GAMA, AR14}, or combining different tracers observed by different surveys \citep{Marin2016, Beutler2016}. These approaches may be subject to limits including a limited relative galaxy bias (samples in a single-tracer survey usually do not differ much in the galaxy bias), and a limited overlapping area (most galaxy surveys are designed to be complementary to each other, in terms of the sky coverage and/or redshift range) \citep{MTreview}.} 

Fortunately, the extended Baryon Oscillation Spectroscopic Survey (eBOSS) project has provided such an opportunity for a proper multi-tracer analysis. Targeted for both Luminous Red Galaxies (LRG) and Emission Line Galaxies (ELG) at $z\in[0.6,1.1]$ in a large overlapping patch of sky, the eBOSS Data Release (DR) 16 provided a total of $\sim550,000$ spectra for the multi-tracer analysis, which is the largest sample for such an analysis to date. This is the natural motivation for this work. In this analysis, we develop new methods for a joint BAO and RSD analysis using the DR16 LRG and ELG sample, and pay particular attention to the mitigation of possible systematics.

The paper is structured as follows. In Section \ref{sec:data}, we describe the observational and simulated datasets used in this analysis, and in Section \ref{sec:method}, we present the method, followed by mock tests and main result of this work in Section \ref{sec:result}, before conclusion and discussion in Section \ref{sec:conclusion}.

This work is one of a series of papers presenting results based on the final eBOSS DR16 samples. The multi-tracer analysis of the same galaxy sample is performed in configuration space to complement this work \citep{Wang2020}. For the LRG sample, produced by \citet{Ross2020}, the correlation function is used to measure BAO and RSD in \citet{Bautista2020}, and the analyses of BAO and RSD from power spectrum are discussed in \citet{GilMarin2020}. The LRG mock challenge for assessing the modelling systematics is described in \citet{Rossi2020}. The ELG catalogues are presented in \citet{Raichoor2020}, and analysed in Fourier space \citep{deMattia2020} and in configuration space \citep{Tamone2020}, respectively. The clustering catalogue of quasar is generated by \citet{Lyke2020,Ross2020}. The quasar mock challenge for assessing the modelling systematics is described in \citet{Smith2020}. The quasar clustering analysis in Fourier space is discussed in \citet{Neveux2020}, and in configuration space in \citet{Hou2020}. Finally, the cosmological implication from the clustering analyses is presented in \citet{eBOSS2020}.

%% file: data.tex
\section{The datasets}
\label{sec:data}

\begin{figure*}
\centering
{\includegraphics[scale=0.8]{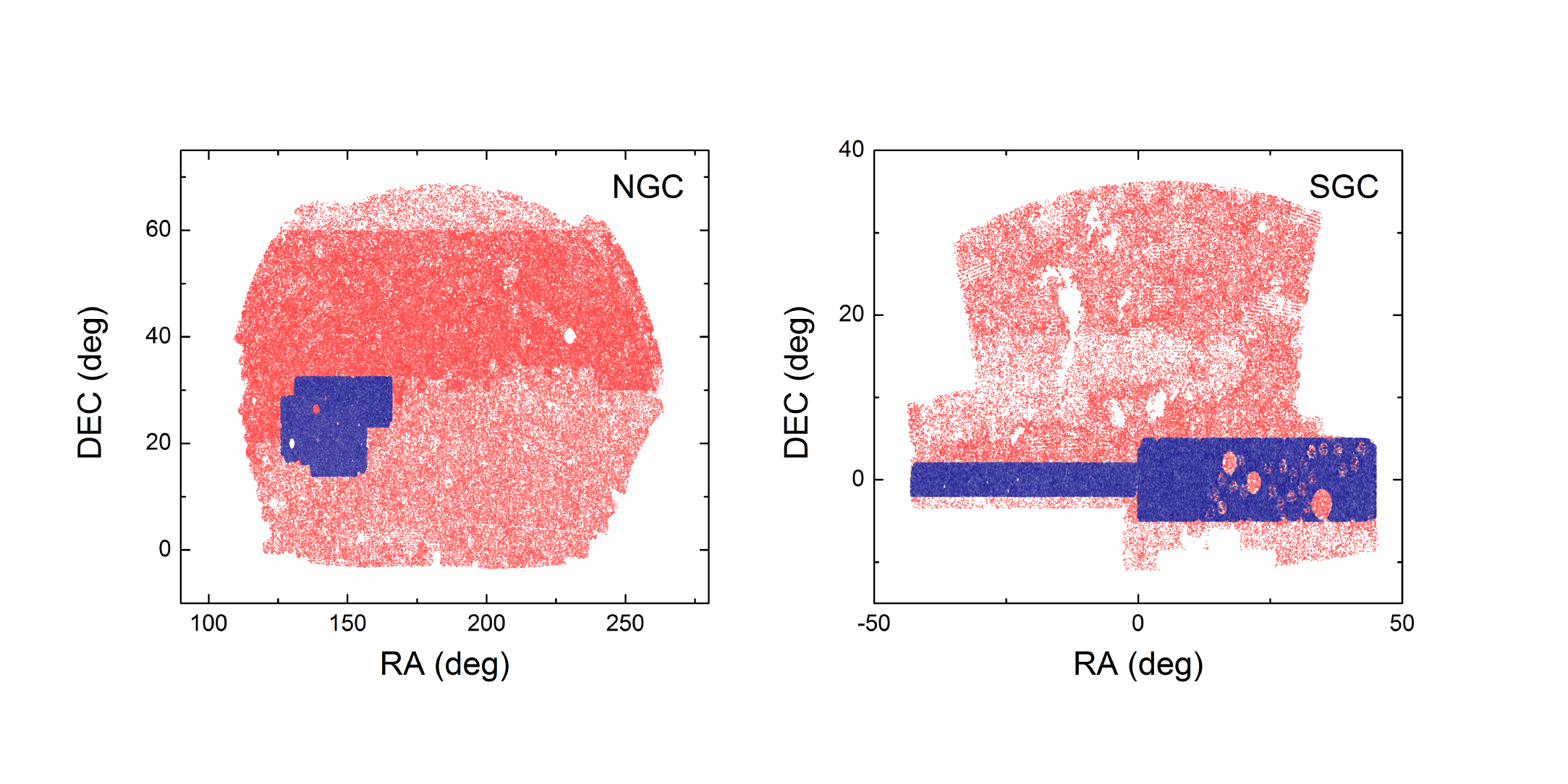}}
\caption{The footprint of the DR16 LRG (larger red region) and ELG (smaller blue) samples in the NGC (left) and SGC (right), respectively, used in this analysis. }
\label{fig:footprint}
\end{figure*}

In this section, we briefly describe the observational and simulated datasets used in this analysis. 

\subsection{The eBOSS DR16 LRG and ELG samples} 

Being part of the Sloan Digital Sky Survey-IV (SDSS-IV) project \citep{sdss4tech}, the eBOSS survey \citep{ebossoverview, eBOSSZhao16} started in 2014 using the 2.5-metre Sloan telescope \citep{sdsstelescope} at the Apache Point Observatory in New Mexico. 

{\tc The LRG targets are selected using optical and infrared imaging data over the entire SDSS imaging footprint. The optical imaging data are taken from the SDSS I/II \citep{sdss12tech} and III \citep{sdss3tech} surveys in five passbands: $u,g,r,i,z$, while the infrared data is provided by the  Wide Field Infrared Survey Explorer (WISE) survey \citep{WISE}. The ELG targets, however, are not selected using the SDSS imaging observations. Instead, the $g,r,z$ bands of the DECam Legacy Survey (DECaLS) \citep{decals} photometric sample is used.} After the eBOSS target selection, which is described in \cite{Ross2020} and \cite{eBOSSELGsel} for the LRG and ELG, respectively, the spectra are taken using the double-armed spectrographs \citep{sdssspectrograph}, which were used for the Baryon Oscillation Spectroscopic Survey (BOSS) mission, as part of the SDSS-III project \citep{sdss3tech}.

The footprint of the LRG and ELG samples is shown in Fig. \ref{fig:footprint}, with statistics in Table \ref{tab:DR16data}. The eBOSS LRG sample used in this work is a combination of the eBOSS LRG with those observed by the BOSS program at $z>0.6$, and it is denoted as `LRGpCMASS' in other companion papers. This sample covers the redshift range of $z\in[0.6,1.0]$ with a sky coverage of $\sim9500 \ {\rm deg}^2$, and consists of approximately $255$ K and $121$ K galaxies in the northern galactic cap (NGC) and southern galactic cap (SGC), respectively. The ELG are selected to cover $z\in[0.6,1.1]$, covering $\sim730 \ {\rm deg}^2$, with $\sim174$ K redshifts in total. 

\begin{figure}
\centering
{\includegraphics[scale=0.28]{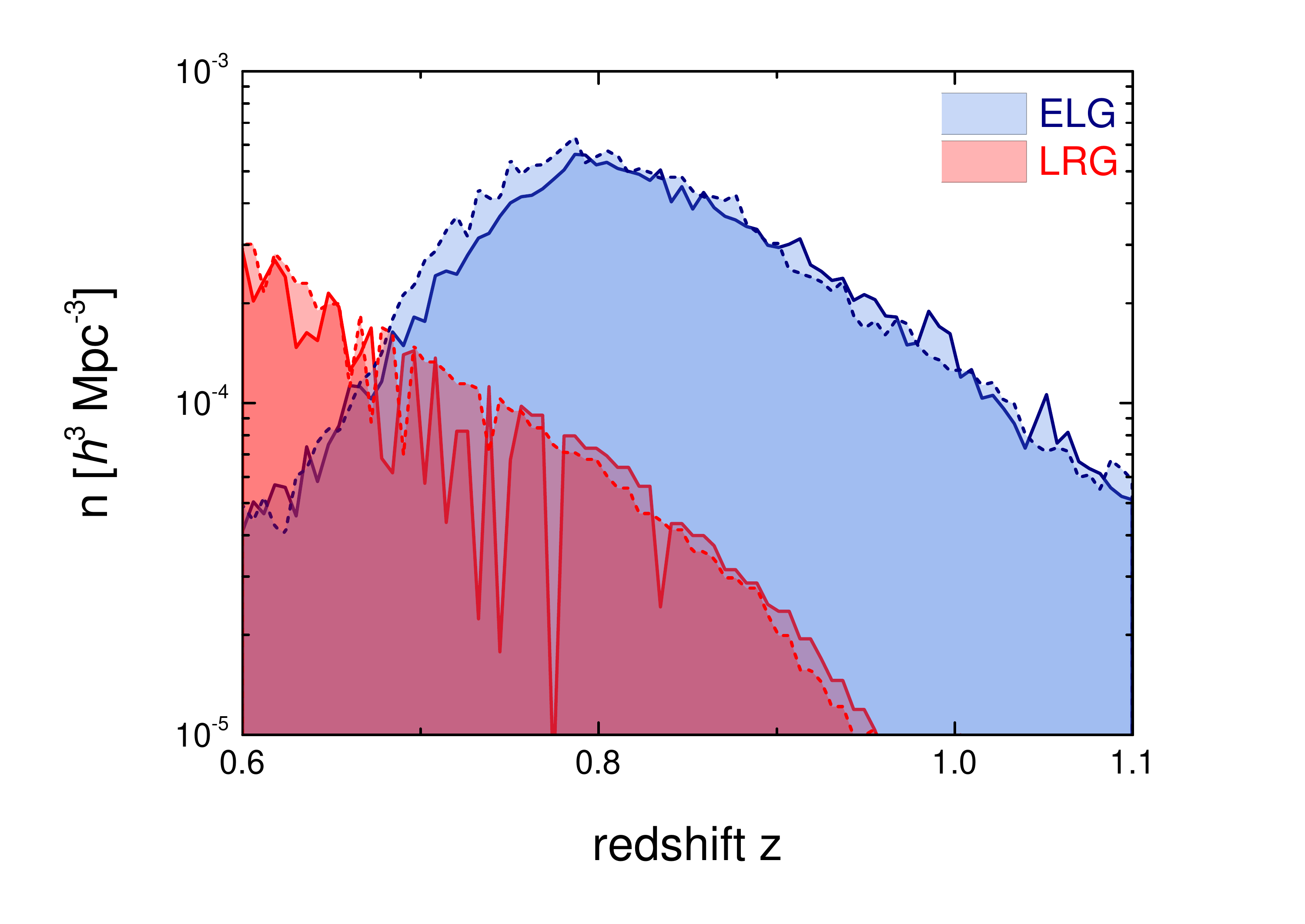}}
\caption{The redshift distribution of the number density of eBOSS DR16 ELG and LRG samples, as illustrated in the legend. For each tracer, the solid and dashed lines represent the NGC and SGC, respectively.}
\label{fig:nz}
\end{figure}

{\tc Fig. \ref{fig:footprint} shows that almost all the ELG are in the footprint of the LRG, but the overlapping region only covers about $8\%$ of the LRG coverage. As we show in a later section (Sec. \ref{sec:result}), this makes the auto-power spectrum of the LRG sample, which is largely dominated by the LRG that do not overlap with the ELG, not closely related to the cross-power between the LRG and ELG samples (a quantitative discussion is in Sec. \ref{sec:result}). The number density distribution in redshift is displayed in Fig. \ref{fig:nz}. Apparently, the overlap between these two samples in redshift is significant, and the densities of both samples are sufficiently high, which enables a multi-tracer exercise.}

\subsection{The simulated mock samples} 

A large number of mock samples, each of which has the same clustering property of the eBOSS DR16 sample, are required to estimate the data covariance matrix. In this analysis, we use the Extended Zel'dovich (EZ) mocks, which consist of $1000$ realisations, produced following the prescription in \cite{EZmock2020,EZmock} {\tc The number of total realisations of the EZmocks we have, which is $2000$ for the LRG and ELG ($1000$ for each), is sufficient given the total number of data points (including those for the cross-power spectrum multipoles) we used, which is $208$, for a joint LRG and ELG analysis \footnote{\tc The Hartlap factor is $0.895$ in our case, which does not significantly deviate from unity, and is included in the likelihood analysis to correct for the data covariance matrix \citep{Hartlap}.}. To reflect the actual situation of the eBOSS observations, observational systematics, including the depth-dependent radial density, angular photometric systematics, fibre collision, redshift failure, \etc, is implemented in the pipeline for producing these mocks (see \citealt{EZmock2020} for more details).} The cosmological parameters used for the EZ mocks are listed in Eq (\ref{eq:pez}), where the parameters are: the physical energy density of cold dark matter and baryons, the sum of neutrino masses, the amplitude of the linear matter power spectrum within $8 h^{-1} \ {\rm Mpc}$, the power index of the primordial power spectrum, and the (derived) scale of the sound horizon at recombination respectively.

\ba {\bf \Theta}&\equiv&\left\{\Omega_c h^2, \Omega_b h^2, \sum M_{\nu}/{\rm eV}, \sigma_8, n_s, \rd/{\rm Mpc} \right\}   \nonumber \\
\label{eq:pfid}&= &\{0.1190,0.022,0,0.8288,0.96,147.74\}|_{\rm f}  \nn \\
\label{eq:pez}&= &\{0.1190,0.022,0,0.8225,0.96,147.66\}|_{\rm EZ}  \ea

We list another set of parameters in Eq (\ref{eq:pfid}), which is the fiducial cosmology we adopt for this analysis \footnote{Throughout the paper, the subscript or superscript `${\rm f}$' denotes the fiducial value.}. 

Note that the EZmocks for different tracers are produced using the same set of random seeds, thus the clustering of different tracers are intrinsically correlated. This is crucial for the multi-tracer analysis in this work.
%and  \cite{Wang2020}, which is a complementary analyse using the same data sample in the configuration space.

\begin{table}
\caption{Abbreviations used in this work with meanings.}
\begin{center}
\begin{tabular}{cc}
\hline\hline
Abbreviation & Meaning \\
\hline
LRG & Luminous Red Galaxies \\
ELG & Emission Line Galaxies \\
LRG (P) & $P_{\ell}(k)$ for LRG \\
LRG (Q) & $Q_{\ell}(k)$ for LRG \\
ELG (P) & $P_{\ell}(k)$ for ELG \\
ELG (Q) & $Q_{\ell}(k)$ for ELG \\
X & The cross power between LRG and ELG \\
QQP &  LRG (Q) + ELG (Q) + X (P)\\
PQP &  LRG (P) + ELG (Q) + X (P)\\
PPP &  LRG (P) + ELG (P) + X (P)\\
$z_{\rm L}$ & $\zeff~(\rm LRG)=0.70$\\ 
$z_{\rm E}$ & $\zeff~(\rm ELG)=0.845$\\ 
$z_{\rm X}$ & $\zeff~(\rm LRG \times ELG)=0.77$\\
FoM & Figure of Merit \\
NGC & Northern Galactic Cap \\
SGC & Southern Galactic Cap \\
LoS & Line of Sight \\
\hline\hline

\end{tabular}

\end{center}
\label{tab:dic}
\end{table}%

\begin{table}
\caption{Statistics of the galaxy sample used in this work. Quantities $P_{\rm shot}$ and $I$, as defined in Eqs (\ref{eq:Phat}) and (\ref{eq:I}), are the shot noise subtracted from the measured monopole, and the normalisation factor for the power spectrum measurement, respectively.}
\begin{adjustbox}{width=\columnwidth,center}
\begin{tabular}{ccccccc}
\hline\hline
                          & LRG(N)        &  LRG(S)      & ELG(N)   & ELG(S)   & X(N)     & X(S) \\
               \hline
Area {\tc $\left({\rm deg}^2\right)$}                 & $6,934$        & $2,560$     & $370$      & $358$      & $370$  & $358$  \\
$N_z$               &  $255,741$   & $121,717$ & $83,769$ & $89,967$ & -           & - \\
$P_{\rm shot} \ {\tc \left(\left[h^{-1} {\rm Mpc}\right]^3\right)}$ &  $12,641$      & $11,995$   & $5,318$   & $4,498$   & -           & - \\
$I$                    & $6.18$          &  $3.00$       & $5.42$     & $5.93$    & $0.88$    & $1.54$   \\
\hline\hline
\end{tabular}
\label{tab:DR16data}
\end{adjustbox}
\end{table}%

%% file: method.tex
\section{Methodology}
\label{sec:method}
We describe the method used in this work, including a brief review of the multi-tracer method, a development of the chained power spectrum to mitigate the angular systematics, and prescriptions of creating the power spectrum template, measuring the power spectrum multipoles with the survey window function, handling the mismatch of $\zeff$ between different tracers, and performing parameter estimations. For the ease of presentation, we include a mini-dictionary in Table \ref{tab:dic} for abbreviations used in this paper.

\subsection{The multi-tracer method}

The clustering of galaxies, as biased tracers of the underlying dark matter field, is subject to the cosmic variance on large scales. The cosmic variance is an intrinsic source of uncertainty for surveys probing a single type of galaxies, but can be significantly suppressed by contrasting the clustering of multiple types of galaxies covering the same range of redshifts and footprints, if the number density of the overlapping tracers are sufficiently high so that the shot noise is negligible on large scales \citep{multi-tracer1,multi-tracer2}.

As described previously, the eBOSS DR16 sample consists of two types of tracers partially overlapping in cosmic volume at $z<1.1$, allowing for a multi-tracer analysis to probe the BAO and RSD jointly.

Under the assumption of Gaussianity, the covariance matrix for power spectrum multipoles of DR16 tracers for a given ${\bf k}$ mode can be modelled as  \citep{FisherRSD},
\ba\label{eq:cov} {\bf C} = \begin{pmatrix}
\LLLL & \LLEE & \LLLE \\
\text{\bf SYM.} &\EEEE  & \EEEL \\
& & \LELE
\end{pmatrix}  \ea where \ba &&  \AAAA =  \left(P_{\rm A}+\frac{1}{n_{\rm A}}\right)^2; \nonumber \\ 
&&\ABAB=\f{1}{2} \left[P_{\rm AB}^2  + \left(P_{\rm A}+\frac{1}{n_{\rm A}}\right)\left(P_{\rm B}+\frac{1}{n_{\rm B}}\right)\right]; \nonumber \\ 
&& \AABB = P_{\rm AB}^2;  \nonumber \\ 
&& \AAAB = P_{\rm AB}\l(P_{\rm A}+\frac{1}{n_{\rm A}}\r),\ea for $\{\A,\B\} \in \{\L,\E \}$. {\tc The auto-power spectrum for tracers A and B are expressed as $P_\A$ and $P_\B$, respectively, and $P_{\A\B}$ denotes the corresponding cross-power. The shot-noise of each tracer are shown as $n_\A$ and $n_\B$, respectively.}

{\tc It is worth noting that using $\bf C$ as the data matrix for the likelihood analysis, or equivalently, using both the auto- and cross-power spectra in the analysis, one essentially measures a ratio between the auto-power spectra of two biased tracers. In the low-noise limit, \ie, $n_\A\rightarrow\infty, \ n_\B\rightarrow\infty$, this ratio can be determined to an infinite accuracy, since the power spectrum for the matter field, which is subject to the cosmic variance, is cancelled out. Interestingly, the RSD parameter, $\beta\equiv f/b$ where $f$ and $b$ are the logarithmic growth rate and the linear bias respectively, is involved in the measured ratio, thus the marginalised uncertainty of the RSD parameter is proportional to the shot noise, \ie, $\beta$ is measured without the cosmic variance \citep{multi-tracer2}. Admittedly, in a realistic situation, the gain from the multi-tracer method can be degraded by a few factors even in the low-noise limit, including the non-Gaussian correction to the distribution of the matter field for example, but this effect is sub-dominant on large scales, on which the modes are more relevant for measuring the RSD.}

\subsection{The effective redshifts}

The measured galaxy cross power spectrum between tracers A and B \footnote{It is the auto power spectrum if A is identical to B.} in a redshift slice is actually a combination of power spectra at multiple redshifts \citep{Zhao19}, \ie, \ba\label{eq:P} P=\frac{\sum P\left(z_{i}\right) \wiA\wiB}{\sum \wiA\wiB}, \ea where $z_i$ is the average redshift for the $i$th galaxy pair made of galaxies $\wiA$ and $\wiB$, and the summation is over all galaxy pairs in the catalog. Traditionally, the clustering analysis is performed at a single effective redshift, $\zeff$, for simplicity. This is an approximation, which can be understood from the following Taylor expansion, \ba\label{eq:Pz} P(z)=P\left(z_{\mathrm{eff}}\right)+P^{\prime}\left(z-z_{\mathrm{eff}}\right)+\frac{1}{2} P^{\prime \prime}\left(z-z_{\mathrm{eff}}\right)^{2}+\mathcal{O}\left(P^{\prime \prime \prime}\right) \ea Combining Eqs. (\ref{eq:P}) and (\ref{eq:Pz}) yields,  \ba P=P\left(z_{\mathrm{eff}}\right)+P^{\prime} \Delta_{1}+\frac{1}{2} P^{\prime \prime} \Delta_{2}+\mathcal{O}\left(P^{\prime \prime \prime}\right) \ea where \ba\label{eq:Delta} \Delta_{1}&=&\frac{\sum z_{i} \wiA\wiB}{\sum \wiA\wiB}-z_{\mathrm{eff}}, \nn\\
\Delta_{2}&=&\frac{\sum z_i^2 \wiA\wiB} {\sum \wiA\wiB}-2 z_{\mathrm{eff}} \frac{\sum z_{i} \wiA\wiB}{\sum \wiA\wiB}+z_{\mathrm{eff}}^{2}. \ea Diminishing $\Delta_1$ by properly defining $\zeff$ as,
\ba\label{eq:zeff} \zeff=\frac{\sum z_{i} \wiA\wiB}{\sum \wiA\wiB}, \ea where $w_i$ is the total weight of each sample, leaves a residual $\Delta_2$ term, \ba \Delta_{2}=\frac{\sum z_{i}^{2} \wiA\wiB}{\sum \wiA\wiB}-\left(\frac{\sum z_{i} \wiA\wiB}{\sum \wiA\wiB}\right)^{2}. \ea Thus one has to make sure that $\Delta_{2}$ (and higher order residuals) is sufficiently small to be ignored for the redshift distribution of the concerning galaxy sample, when using a fixed power spectrum template, otherwise the analysis may be subject to systematics. We explicitly evaluate $\Delta$ defined in Eq. (\ref{eq:Delta}) at $\zeff$ shown in Eq. (\ref{eq:zeff}) for our samples, and summarise the result in Table \ref{tab:Delta}. By construction, $\Delta_1$ vanishes for each power spectrum at its own effective redshift, \eg, $\zeff=0.700, 0.845$ and $0.770$ for $\PLL$, $\PEE$ and $\PLE$, respectively, and $\Delta_2$ gets minimised in this case. We have numerically confirmed that, in this case, the second-order correction term, $P''\Delta_2/2$, is safely negligible compared to the leading term \footnote{\tc In order to compare the correction term to the leading term, we in practice evaluate $P'$ and $P''$ numerically using a three-point finite difference scheme at the fiducial cosmology.}. However, this may not hold if the analysis is performed at a redshift that is significantly different from the effective redshift. For example, analysing the LRG sample at $z=0.845$ would require $\Delta_2=0.028$ to compensate, which is four times larger than that at its own $\zeff$, thus the second or higher order correction terms may have to be included in the template to avoid theoretical systematics.

\begin{table}
\caption{The quantities $\Delta_1$ and $\Delta_2$ for various power spectrum types at different effective redshifts.}
\begin{center}
\begin{tabular}{cccccccccc}
\hline\hline
&  \multicolumn{2}{c}{$\PLL$} & \multicolumn{2}{c}{$\PEE$} &  \multicolumn{2}{c}{$\PLE$} \\
\hline
$\zeff$ & $\Delta_1$ & $\Delta_2$ & $\Delta_1$ & $\Delta_2$ & $\Delta_1$ & $\Delta_2$ \\
\hline
$0.700$ & $0$ & $0.007$ & $0.145$ & $0.031$  & $0.070$ & $0.012$ \\
$0.845$ & $-0.145$ & $0.028$ & $0$ & $0.011$  & $-0.075$ & $0.013$ \\
$0.770$ & $-0.070$ & $0.012$ & $0.075$ & $0.017$  & $0$ & $0.007$ \\
\hline\hline
\end{tabular}
\end{center}
\label{tab:Delta}
\end{table}%

\subsection{The time dependence of the BAO and RSD parameters}

Care must be taken when cross-correlating galaxy samples, because different samples may have different effective redshifts, even if they perfectly overlap. One could, in principle, use different $\zeff$ to generate templates for auto-correlation of each tracer, and for their cross-correlation respectively, but this inevitably requires additional parameters for BAO and RSD, which may degrade the efficiency of the multi-tracer technique. One way out is to relate the BAO and RSD parameters at different redshifts by a general parametrisation. For this purpose, we follow \cite{Zhao19} to use the parametrisation for evaluating the optimal redshift weights, when necessary.

\ba\label{eq:BAORSDz} \alpha_{\bot}(z)&=& \alpha_{\bot}(\zp)+\left[\alpha_{\|}(\zp) - \alpha_{\bot}(\zp)  \right]x, \nn\\ 
\alpha_{\|}(z)&=& \alpha_{\|}(\zp)+2\left[\alpha_{\|}(\zp) - \alpha_{\bot}(\zp)  \right]x, \nn\\
f(z)&=&f(\zp)\left( \frac{1+z}{1+\zp}  \right)^{3\gamma} \left[\frac{\alpha_{\|}(z)}  {\apar(\zp)}   \frac{H_{\rm f}(\zp)} {H_{\rm f}(z)}\right]^{2\gamma},
\ea where $\gamma$ is the growth index introduced in \cite{Lindergamma}, $\zp$ is the pivot redshift, $x\equiv\chi_{\rm f}(z)/\chi_{\rm f}(z_{\rm p})-1$ and $\chi(z)$ and $H(z)$ are the comoving distance and the Hubble function at redshift $z$, respectively \footnote{\tc The pivot redshift $\zp$ defines a redshift at which the Taylor expansion is performed, \ie, $x(\zp)=0$, thus $\zp$ is usually chosen so that $x$ remains small in the redshift range of interest. A convenient choice of $\zp$ is the effective redshift of a galaxy sample, which is adopted in this work.}. This set of parametrisation has been proven to be sufficiently general to cover a broad class of cosmologies in a wide redshift range \citep{ZPW,SoL}. In this work, we use this framework to relate BAO and RSD parameters at $z=0.77$ and $z=0.845$, which is well within the validity of this parametrisation, given the uncertainty of the eBOSS DR16 sample.

\subsection{Measuring the auto and cross power spectrum multipoles}
\label{sec:pkmeasure}

The measurement of the power spectrum multipoles can be performed efficiently using the Fast Fourier Transformation (FFT) \citep{pkFFT,pkFFT2}, based on the Yamamoto estimator \citep{Yama2006},

\ba\label{eq:Phat}  \hat{P}_{\ell}(k)&=& \frac{2 \ell+1}{I} \int \frac{\mathrm{d} \Omega_{k}}{4 \pi}\left[\int \mathrm{d} \boldsymbol{r}_{1} F\left(\boldsymbol{r}_{1}\right) \mathrm{e}^{\mathrm{i} \boldsymbol{k} \cdot \boldsymbol{r}_{1}}\right. \nn \\ &&\left.\times \int \mathrm{d} \boldsymbol{r}_{2} F\left(\boldsymbol{r}_{2}\right) \mathrm{e}^{-\mathrm{i} \boldsymbol{k} \cdot \boldsymbol{r}_{2}} \mathcal{L}_{\ell}\left(\hat{\boldsymbol{k}} \cdot \hat{\boldsymbol{r}}_{2}\right)-P_{\rm shot}\right],  \ea {\tc where $P_{\rm shot}$ is the shot noise component, and the intergral is over the entire volume of the survey.} The line-of-sight (LOS) of pairs is approximated as the LOS of one of the galaxies in the pair, \ie, $\mathcal{L}_{\ell}(\hat{\boldsymbol{k}} \cdot \hat{\boldsymbol{r}}) \simeq \mathcal{L}_{\ell}(\hat{\boldsymbol{k}} \cdot \hat{\boldsymbol{r}}_{2})$, and the overdensity field is estimated as \cite{FKP},
\ba F(\boldsymbol{r})=\frac{w(\boldsymbol{r})}{I^{1 / 2}}\left[n(\boldsymbol{r})-\alpha n_{\mathrm{s}}(\boldsymbol{r})\right], \ea where $w$ is the total weight of each galaxy, and $n, n_s$ denotes the number density of the data and random samples, respectively. The quantity $\alpha$ is the ratio of the weighted numbers of the data and random, and the normalisation $I$ is evaluated as, \ba\label{eq:I} I \equiv \int \mathrm{d} \boldsymbol{r} \ w^2(\boldsymbol{r}) n^{2}(\boldsymbol{r})\simeq \alpha\sum_i w_i^2 n_{s,i} \ea Note that the above approximation using sums over the randoms is only valid for the auto-power. For the cross power, one has to take the overlapping geometry into account. A practical way is to assign random galaxies of both tracers onto a grid, and for each tracer, compute $w^2 n$ for each grid cell, and compute the product $\sqrt{(w^2 \ n)_{\rm A}}\sqrt{(w^2 \ n)_{\rm B}}$ for each grid cell, and sum over the cells. The final result for $I$ and $P_{\rm shot}$ for each tracer is summarised in Table \ref{tab:DR16data}.

We use a $1024^3$ grid for evaluating $F$ and $w^2n$, use a fourth-order B-spline for interpolation, and correct for the aliasing effect following \cite{Jing05}. We use the following estimator to measure the cross power between tracers A and B, which makes use of the {\tc spherical harmonic Addition Theorem} \citep{AT} to factorise the Legendre polynomial into a product of spherical harmonics,

\ba \widehat{P}_{\ell}(k)=\frac{2 \ell+1}{2I} \int \frac{\mathrm{d} \Omega_{k}}{4 \pi} \l[F_{0,{\rm A}}(\mathbf{k}) F_{\ell,{\rm B}}(-\mathbf{k}) + F_{0,{\rm B}}(\mathbf{k}) F_{\ell,{\rm A}}(-\mathbf{k})\r],\nn  \ea where 

\ba \begin{aligned} F_{\ell}(\mathbf{k}) & \equiv \int \operatorname{dr} F(\mathbf{r}) e^{i \mathbf{k} \cdot \mathbf{r}} \mathcal{L}_{\ell}(\hat{\mathbf{k}} \cdot \hat{\mathbf{r}}) \\ &=\frac{4 \pi}{2 \ell+1} \sum_{m=-\ell}^{\ell} Y_{\ell m}(\hat{\mathbf{k}}) \int \operatorname{dr} F(\mathbf{r}) Y_{\ell m}^{*}(\hat{\mathbf{r}}) e^{i \mathbf{k} \cdot \mathbf{r}} \end{aligned} \nn \ea

\subsection{The chained power spectrum multipoles}

To minimize the impact from unknown systematics, we propose a new observable to use, which is the ``chained power spectrum multipoles'', as defined below, which is immune to any angular systematics, \ie, any contaminant coupling to the transverse mode.

\begin{figure}
\centering
{\includegraphics[scale=0.32]{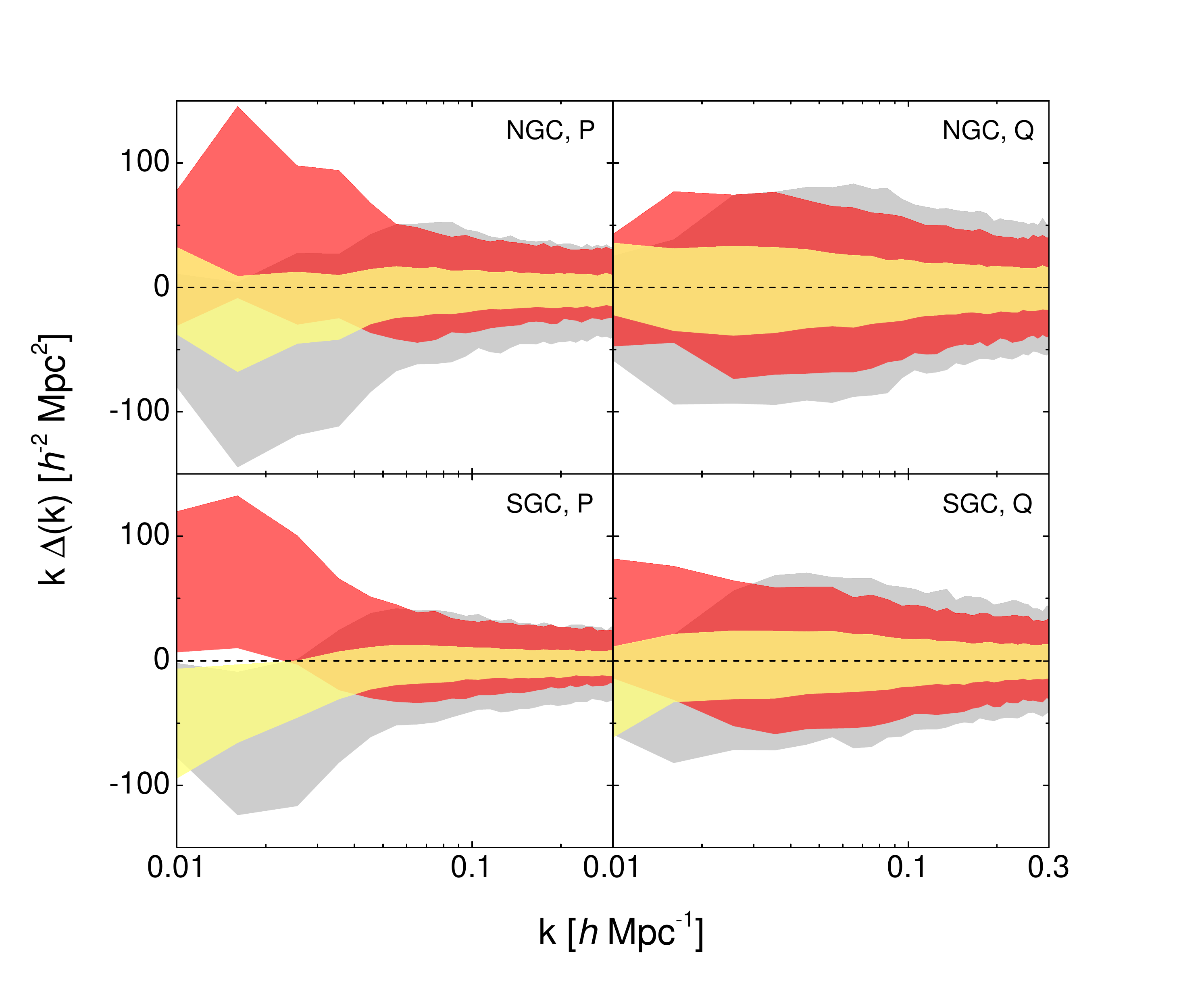}}
\caption{A demonstration of mitigating the angular systematics using the chained power spectrum multipoles. The quantity $\Delta(k)$ shows the difference in multipoles of the power spectrum $P_{\ell}(k)$ (left panels) or the chained power spectrum $Q_{\ell}(k)$ (right panels) measured from the contaminated or uncontaminated versions of EZmocks of the ELG sample. The filled bands show the 68\% CL range of $k\Delta(k)$ for the monopole (bottom gray layer), quadrupole (middle red) and hexadecapole (top yellow). The dashed line shows $\Delta(k)=0$ as a reference.}
\label{fig:kDk}
\end{figure}

The observed power spectra $P^{\rm obs}(k,\mu)$ can be understood as follows. {\tc If the angular systematics $X(k)$ only contaminates the transverse mode, \ie, the $\mu=0$ mode, then it can be modeled in the following way, as discussed in \citep{Hand17}}\footnote{\tc It is true that this is a toy model for the angular systematics, but this captures the primary systematics in the eBOSS ELG sample, as we demonstrate in the mock test.}, 
\ba\label{eq:Pkmuobs} P^{\rm obs}(k,\mu) =  P^{\rm true}(k,\mu) + X(k)\delta_{\rm D}(\mu),\ea where $\delta_{\rm D}$ is the Dirac-$\delta$ function. A multipole expansion of Eq. (\ref{eq:Pkmuobs}) shows, \ba\label{eq:Pellobs} P_{\ell}^{\rm obs}(k) =  P_{\ell}^{\rm true}(k) + \frac{2\ell+1}{2}X(k)\mathcal{L}_{\ell}(0), \ea with $\mathcal{L}_{\ell}$ being the Legendre polynomial of order $\ell$. Proceed Eq. (\ref{eq:Pellobs}) to the next non-vanishing order, we get, \ba\label{eq:Pellp2obs} P_{\ell+2}^{\rm obs}(k) =  P_{\ell+2}^{\rm true}(k) + \frac{2\ell+5}{2}X(k)\mathcal{L}_{\ell+2}(0). \ea Eliminating $X(k)$ from Eqs. (\ref{eq:Pellobs}) and (\ref{eq:Pellp2obs}), we obtain the following relation, \ba\label{eq:QQ} Q_{\ell}^{\rm obs} = Q_{\ell}^{\rm true}, \ea where $Q_{\ell}$ is the chained power spectrum multipoles,
\ba\label{eq:Qell} Q_{\ell} \equiv P_{\ell} - A_{\ell} P_{\ell+2}, \ea and \ba A_{\ell} \equiv \frac{(2\ell+1)\mathcal{L}_{\ell}(0)}{(2\ell+5)\mathcal{L}_{\ell+2}(0)}. \ea Unlike the observed $P_{\ell}$, the observed $Q_{\ell}$ is immune to the angular systematics, as demonstrated by Eq. (\ref{eq:QQ}), thus is a better quantity to use for data analysis. 

For the first three multipoles of $Q$, Eq. (\ref{eq:Qell}) means,\ba\label{eq:matrix}  
\begin{pmatrix}
Q_0  \\
Q_2  \\
Q_4 
\end{pmatrix}  = \begin{pmatrix}
1 & -A_0 & 0 &0  \\
0 & 1 & -A_2 & 0  \\
0 & 0 & 1 & -A_4 
\end{pmatrix}  \begin{pmatrix}
P_0  \\
P_2  \\
P_4  \\ 
P_6 
\end{pmatrix}
\ea To reconstruct $P$ from $Q$, a truncation in $P_{\ell}$ is necessary, otherwise the above matrix equation is not invertible. As $P_{\ell}=0 \ (\ell>4)$ in linear theory, we show an example in which $P_6$ is set to zero after finding $Q_4$ from data. An matrix inversion of the first $3\times3$ block of the transformation matrix in Eq. (\ref{eq:matrix}) yields the cleaned $P$, denoted as $P^{\rm c}$, \ba\label{eq:matrixinv}  
\begin{pmatrix}
P^{\rm c}_0  \\
P^{\rm c}_2  \\
P^{\rm c}_4 
\end{pmatrix}  = \begin{pmatrix}
1 & A_0 & A_0 A_2   \\
0 & 1 & A_2  \\
0 & 0 & 1  
\end{pmatrix}  \begin{pmatrix}
Q_0  \\
Q_2  \\
Q_4 
\end{pmatrix} = \begin{pmatrix}
1 & 0 & 0 &  -A_0A_2A_4   \\
0 & 1 & 0 & -A_2A_4  \\
0 & 0 & 1 & -A_4 
\end{pmatrix}  \begin{pmatrix}
P_0  \\
P_2  \\
P_4  \\
P_6
\end{pmatrix}. 
\ea This equation is physically transparent: the role of measured $P_6$, which is supposed to be zero as a theoretical prior chosen in this example, is to provide an estimate of the transverse contamination, $X(k)$.

We caution that the window function of galaxy surveys can complicate the above formalism, because $P_{\ell}^{\rm obs}$ receives contributions from not only $P_{\ell}^{\rm true}$, but also $P_{\ell'}^{\rm true}$, which are the true $P(k)$ multipoles with similar orders, due to the convolution with the anisotropic survey window function. A complete prescription for mitigating the angular systematics with the window function effect is beyond the scope of this paper, but we argue that the chained power spectrum method developed here can remove the primary angular systematics, because $P_{\ell}^{\rm true}$ dominates $P_{\ell}^{\rm obs}$, even with the window function effect. 

For the eBOSS DR16 sample, we find that $Q_4$ is rather noisy, as it involves the $P_6$ component, which is barely informative on linear scales. We thus choose not to use $Q_4$ for this work. Admittedly, we learn less of the galaxy clustering from $Q_0$ and $Q_2$ than from $P_0, P_2$ and $P_4$, but the information loss can be largely compensated by adding $\PlX$, multipoles of cross power spectrum between LRG and ELG, to the analysis. As LRG and ELG are selected using different photometry, we assume that the angular systematics of these tracers are uncorrelated, \ie, $\PlX$ is immune to angular systematics. 

In principle, we can use the following data vectors for analysis,
 \ba\label{eq:PQP} {\rm PPP}&\equiv&\l(P^{\rm L}_0, P^{\rm L}_2, P^{\rm L}_4, P^{\rm E}_0, P^{\rm E}_2, P^{\rm E}_4, P^{\rm X}_0, P^{\rm X}_2, P^{\rm X}_4 \r)^T; \nn \\ 
{\rm PQP}&\equiv&\l(P^{\rm L}_0, P^{\rm L}_2, P^{\rm L}_4, Q^{\rm E}_0, Q^{\rm E}_2, P^{\rm X}_0, P^{\rm X}_2, P^{\rm X}_4 \r)^T; \nn \\
{\rm QQP}&\equiv&\l(Q^{\rm L}_0, Q^{\rm L}_2, Q^{\rm E}_0, Q^{\rm E}_2, P^{\rm X}_0, P^{\rm X}_2, P^{\rm X}_4 \r)^T, \ea where `L', `E' and `X' denote observables for the LRG, ELG and their cross correlation, respectively. Apparently, PPP and QQP are the most aggressive and most conservative combinations, respectively, and PQP is in between. We shall make the choice in Sec. \ref{sec:result}, after validating our pipeline by performing analyses on the mocks using all these combinations.

\subsection{The power spectrum template}

The TNS model \citep{TNS} has been widely used as a theoretical template for analyses using the auto-power spectrum with the linear and nonlocal bias terms included \citep{MR09,FBRSD16}. For multiple tracers, the TNS model can be generalised as follows, 
\ba
\label{eq:Pgkmu}  P_{\rm g}^{\rm AB}(k,\mu) &=& {D_{\rm FoG}}\left(k,\mu\right)  \left[P_{\rm g,\dd}^{\rm AB} (k)\right. \nonumber \\
&&+  2f\mu^2P_{\rm g,\dt}^{\rm AB} (k)  + f^2\mu^4 P_{\rm\vv}^{\rm AB}(k)  \nonumber \\
&& \left.+ A^{\rm AB}(k,\mu)+B_{}^{\rm AB}(k,\mu)\right],\ea
where 
\ba
\label{eq:Pgdd} P_{\rm g,\dd}^{\rm AB}(k) &=& \bIA\bIB P_{\dd}(k)+\l(\bIA\bIIB+\bIB\bIIA\r) P_{\rm b2,\updelta}(k)  \nn \\ 
&& +\l(\bsA\bIB+\bsB\bIA \r)P_{\rm bs2,\updelta}(k)   \nn \\
&& + \l( \bsA\bIIB+\bsB\bIIA \r)P_{\rm b2s2}(k) \nn \\
&& + \l(\bIIIA\bIB+\bIIIB\bIA \r) \sigma_3^2(k)P_m^{\rm L}(k)  \nn \\
&& + \bIIA\bIIB  P_{\rm b22}(k) + \bsA\bsB P_{\rm bs22}(k) + N_{\rm AB}, \nn  \\ \\ 
P_{\rm g,\dt}^{\rm AB}(k) &=& \f{1}{2} \l[\l(\bIA+\bIB\r) P_{\dt}(k)+\l(\bIIA+\bIIB\r)P_{\rm b2,\uptheta}(k)\r.\nonumber \\
&& + \l(\bsA+\bsB\r) P_{\rm bs2,\uptheta}(k) \nn \\
&& + \l.\l(\bIIIA+\bIIIB\r) \sigma_3^2(k)P_m^{\rm L}(k)\r], \\ \nn \\ 
P_{\rm g,\vv}(k) &=&P_{\vv}(k),\\
D_{\rm FoG} (k,\mu)&=& \left\{1+\left[k\mu\sigma_v\right]^2/2\right\}^{-2}, \ea with a full derivation of the $A^{\rm AB}$ and $B^{\rm AB}$ terms for the multi-tracer case included in Appendix \ref{sec:xTNS} \footnote{The numeric code for evaluating the $A^{\rm AB},B^{\rm AB}$ terms for the cross power is avaiable at \url{http://www2.yukawa.kyoto-u.ac.jp/~atsushi.taruya/cpt_pack.html}.}. This template restores the form for the auto-power if ${\rm A}={\rm B}$.

The subscripts $\delta$ and $\theta$ denote the overdensity and velocity divergence fields, respectively, and $P_{\dd}, P_{\dt}$ and $P_{\vv}$ are the corresponding nonlinear auto- or cross-power spectrum, evaluated using the regularised perturbation theory (RegPT) up to second order \citep{RegPT}. The linear matter power spectrum $P_m^{\rm L}$ is calculated using {\tt CAMB} \citep{CAMB}. Terms $b_1$ and $b_2$ stand for the linear bias and the second-order local bias respectively. We have eliminated the second-order non-local bias $b_{\rm s2}$ and the third-order non-local bias $b_{\rm 3nl}$ using the following relation \citep{nonlocalb1,nonlocalb2,nonlocalb3},
\ba 
\label{eq:bs2b3nl} &&b_{\rm s2} = -\frac{4}{7}\left(b_1-1\right),\nonumber \\
&&b_{\rm 3nl} = \frac{32}{315}\left(b_1-1\right).
\ea

Note that the template of the cross power cannot be represented using that for the auto-power by redefining {\tc a new} set of bias parameters in the framework of the TNS model, as explicitly shown in Appendix \ref{sec:auto-x}, therefore we choose not to introduce an additional set of bias parameters for the cross power for theoretical consistency, although this approach is taken for the analysis in the configuration space \citep{AR14,Wang2020}.

\subsection {The Alcock-Paczynski effect}

The Alcock-Paczynski (AP) effect \citep{AP} distorts the observed power spectrum due to a possible mismatch between the input cosmology, which is used to convert redshifts to distances, and the true cosmology hidden in the observations. This effect creates anisotropy at the background level, via the following dilation parameters,
\ba \abot = \f{\DM(z) \rd^{\rm f}}{\DM^{\rm f}(z) \rd}; \ \ \apar = \f{\DH(z) \rd^{\rm f}}{\DH^{\rm f}(z) \rd}, \ea with \ba \DM(z) = (1+z)D_{\rm A}(z); \ \ \DH(z) = c/H(z), \ea where $D_{\rm A}(z), H(z)$ are the angular diameter distance and the Hubble function at redshift $z$, respectively, and $c$ is the speed of light. The $\alpha$ parameters then distort the wavenumber $k$ and $\mu$, which is the cosine of the angle between the LoS and the galaxy pair, in the following way,
\ba k^{\prime}=\frac{k}{\alpha_{\perp}}\left[1+\mu^{2}\left(\frac{1}{F^{2}}-1\right)\right]^{1 / 2}; \ \mu^{\prime}=\frac{\mu}{F}\left[1+\mu^{2}\left(\frac{1}{F^{2}}-1\right)\right]^{-1 / 2}, \nn \ea where $F=\apar/\abot$ \citep{BPH96}, and the resultant power spectrum multipole with order $\ell$ reads,
\ba P_{\ell}^{\rm AB}(k)= \frac{(2 \ell+1)}{2 \alpha_{\perp}^{2} \alpha_{\|}} \int_{-1}^{1} \d \mu \  P^{\rm AB}_{\mathrm{g}}\left[k^{\prime}(k, \mu), \mu^{\prime}(\mu)\right] \mathcal{L}_{\ell}(\mu). \ea

\subsection{The survey window function}

To account for the geometry of the survey,  we follow \cite{pkmask} to compute the survey window functions for the auto-power spectrum of all tracers, and the cross-power spectrum between LRG and ELG, using the pair-count approach,

\ba W^{\rm AB}_{\ell}(s)=\frac{(2 \ell+1)}{I \alpha^{-2}} \sum_{i, j}^{N_{\mathrm{ran}}} \frac{w^{\rm A}_{\mathrm{tot}}\left(\mathbf{x}_{i}\right) w^{\rm B}_{\mathrm{tot}}\left(\mathbf{x}_{j}+\mathbf{s}\right)}{4 \pi s^{3} \Delta ({\rm log} \ s)} \mathcal{L}_{\ell}\left(\hat{\mathbf{x}}_{\mathrm{los}} \cdot \hat{\mathbf{s}}\right), \ea where superscripts $A, B$ denote different types of tracers, and again, $A=B$ is the limit for the auto-correlation. This is a multi-tracer generalisation of the formalism in \cite{GilMarin2020}, and note that, the factor $I$ appears in the denominator, as suggested by \cite{RIC}, to match the normalisation in the measurement of power spectrum, so that the final BAO and RSD measurement does not depend on how exactly the power spectrum is normalised. 

\subsection{The radial integral constraint} 
Due to the ignorance of the true selection function of the galaxy survey, {\tc which is needed for a clustering analysis,} the redshift distribution of actual observations, $n(z)$, is used instead {\tc as the selection function for analysis. This can in principle bias the final measurement of BAO and RSD parameters if not accounted for. The resultant bias, quantified as the radial integral constraint (RIC), is recently investigated in \cite{RIC}}, and corrected for {\tc in the theoretical model for the power spectrum in} the eBOSS DR16 ELG analysis \citep{deMattia2020}. In this work, we take a different approach, namely, subtracting the RIC component from the data directly. In practice, we analyse two sets of EZmocks, with different treatments of the randoms so that one has the RIC effect, while the other does not. Then a comparison of the measured power spectrum from these two sets provides an estimate of the RIC component of the power spectrum. {\tc The two approaches are indistinguishable if the RIC barely depends on cosmology, which is proven to be true for the DR16 sample using mock.} 

\subsection{Parameter estimation}
\label{sec:parameter}

With a modified version of {\tt CosmoMC} \citep{cosmomc} {\tc which supports sampling the BAO and RSD parameters using a TNS template}, we use the Markov Chain Monte Carlo (MCMC) algorithm to sample the following general parameter space, \ba {\bf P}=\l\{\abot,\apar,\fs,\l\{b_1\sigma_8\r\},\l\{b_2\sigma_8\r\},\l\{\sigma_v\r\}, \l\{ N\r\} \r\}, \ea where quantities in the inner bracket denote a collection of parameters for each tracer in each galactic cap, \eg, \ba \l\{b_1\sigma_8\r\} = \l\{\bsILN,\bsILS, \bsIEN,\bsIES \r\}, \nn \ea and $N$ is fixed to zero for the cross power. We list wide flat priors for these parameters in Table \ref{tab:prior}.

{\tc Note that we use separate sets of bias parameters for the NGC and SGC, to account for the fact that unknown systematics may yield slightly different amplitudes of power spectra in different galactic caps. This treatment is consistent with BOSS DR12 analyses \citep{alam}, and with other eBOSS DR16 analyses \citep{GilMarin2020,deMattia2020}.}

\begin{table}
\caption{\tc Parameters sampled with flat priors used in this analysis.}
\begin{center}
\begin{tabular}{cc}
\hline\hline
Parameter & Flat prior \\
\hline
$\abot$ & $[0.5,1.5]$\\
$\apar$ & $[0.5,1.5]$\\
$\fs$ & $[0,3]$\\
$b_1\sigma_8$ & $[0,10]$\\
$b_2\sigma_8$ & $[-10,10]$\\
$\sigma_v$ & $[0,20]$\\
$N$ & $[-5,5]\times \ P_{\rm shot}$\\
\hline\hline
\end{tabular}
\end{center}
\label{tab:prior}
\end{table}%

By default, we assign a full set of the above parameters for a joint analysis at three effective redshift, resulting in joint BAO and RSD measurements at $\zL=0.70,\zX=0.77$ and $\zE=0.845$, dubbed as the `$3z$' measurement. Alternatively, we also perform a `$2z$' measurement, by relating parameters at $z_{\rm X}$ with those at $z_{\rm E}$ using the parametrisation introduced in Eq. (\ref{eq:BAORSDz}) with $\zp=\zE=0.845$. This essentially spends the information of the cross power for measuring parameters at the effective redshift of the ELG sample, which yields a joint measurement at $\zL$ and $\zE$. The reason for combining the autopower of the ELG with the cross power is the following,
\begin{itemize}
    \item  As we shall present in Sec. \ref{sec:result}, the power spectrum of the ELG, or the cross power on their own, struggles to constrain the BAO parameter well {\tc due to the low signal-to-noise ratio}, which results in loose and highly non-Gaussian constraints without combining with each other;
    \item The ELG sample is known to be much more contaminated by the systematics than the LRG sample \citep{deMattia2020,Tamone2020}, thus combining with the cross power is an efficient way to mitigate the systematics, in addition to using the $Q_{\ell}$'s as observables;
    \item The LRG sample, on the other hand, is much less subjective to systematics \citep{GilMarin2020,Bautista2020}, and it can provide a decent measurement on its own, making it unnecessary to combine it with the cross power;
    \item Tomographic information on the lightcone is key for probing physics including the nature of dark energy \citep{ZhaoDE17}, thus we choose not to compress all the power spectra into a measurement at a single redshift.
\end{itemize}

All the above arguments support for performing the `$2z$' measurement, which will be presented in Sec. \ref{sec:result} as the primary result of this paper. 

In this work, we use data points in the range of $k\in[0.02,0.15] \ \kunit$ for all spectra, as motivated by the LRG analysis \citep{GilMarin2020}, and have confirmed that this is an appropriate choice for the multi-tracer analysis, based on analyses using the mocks. In all cases, we combine the likelihoods for the NGC and SGC using a direct sum, and properly correct for the (inverse) data covariance matrix with relevant correction factors suggested by \cite{Hartlap,Percival14}.

We analyse the chains using {\tt GetDist} \citep{getdist}, after the chains are fully converged, {\tc namely, the Gelman and Rubin statistics $R-1<0.01$ in all cases \citep{R-1,cosmomc}.}

\begin{figure*}
\centering
{\includegraphics[scale=0.45]{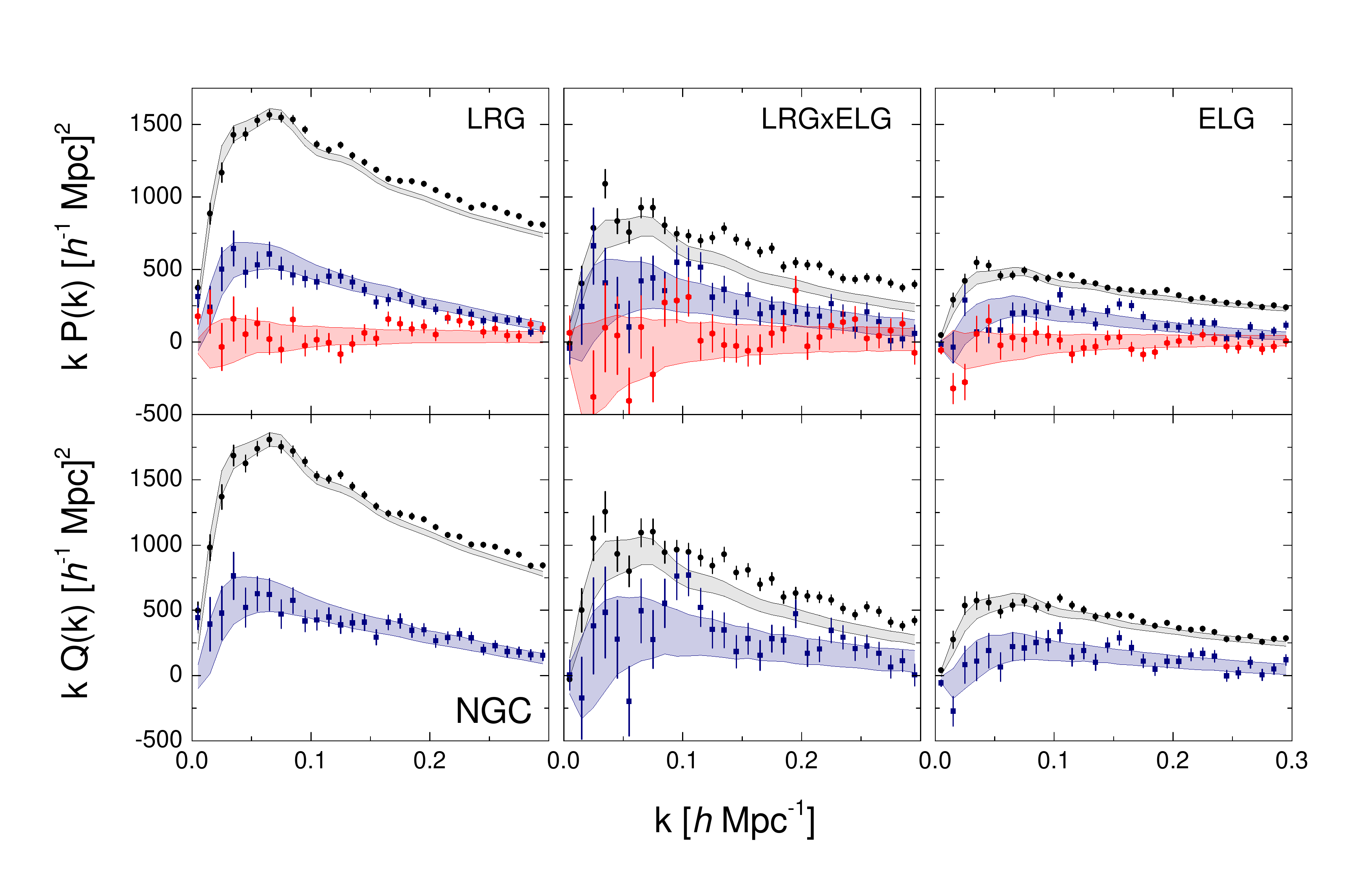}}
\caption{The power spectrum multipoles measured from $1000$ realisations of the EZmocks (filled bands) and from the DR16 data (data points with error bars) for the LRG (left), ELG (right) and the cross correlation between LRG and ELG (middle). }
\label{fig:PQ_NGC}
\end{figure*}

\begin{figure*}
\centering
{\includegraphics[scale=0.45]{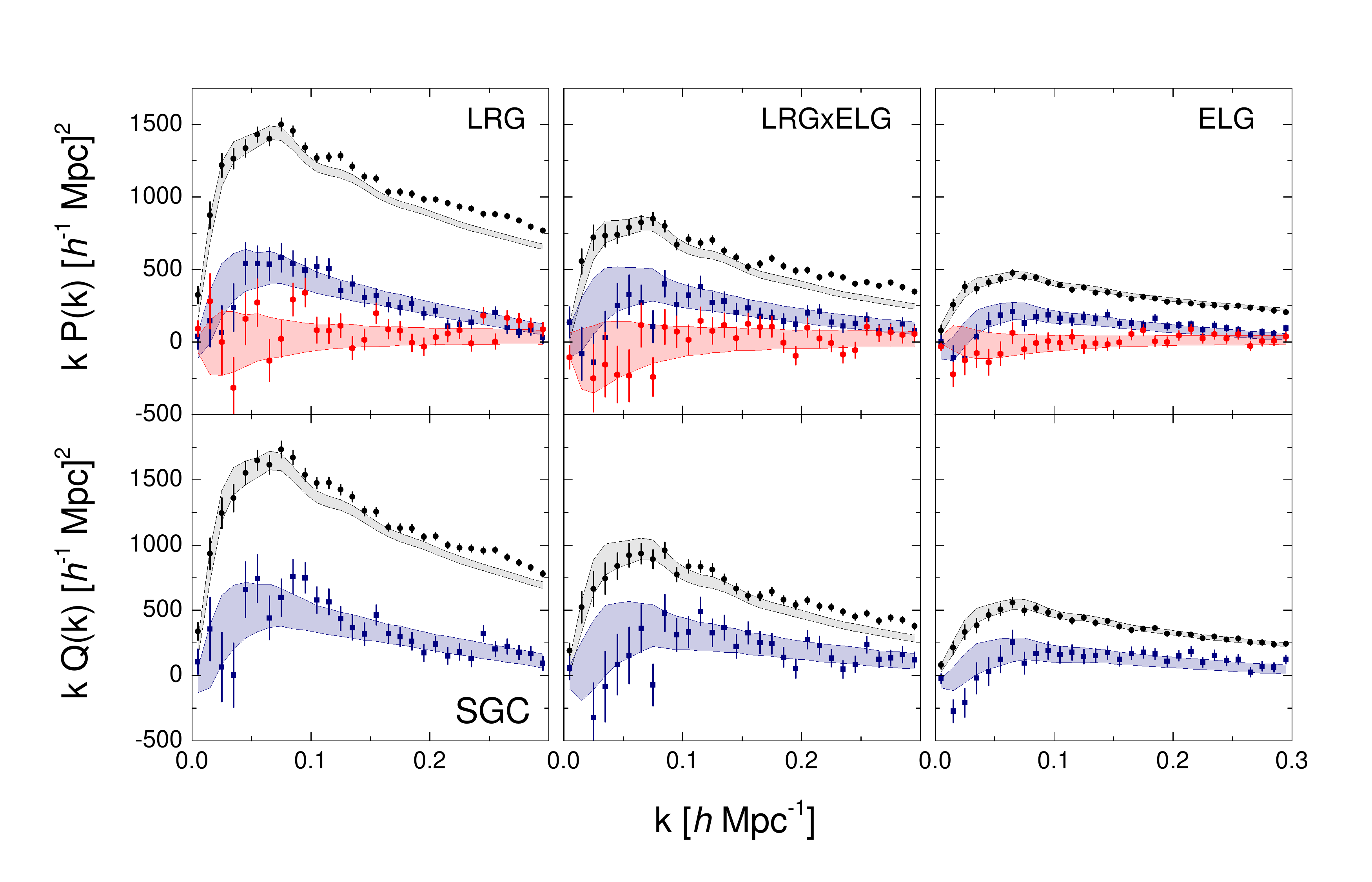}}
\caption{Same as Fig. \ref{fig:PQ_NGC}, but for the SGC.}
\label{fig:PQ_SGC}
\end{figure*}

\begin{figure*}
\centering
{\includegraphics[scale=0.35]{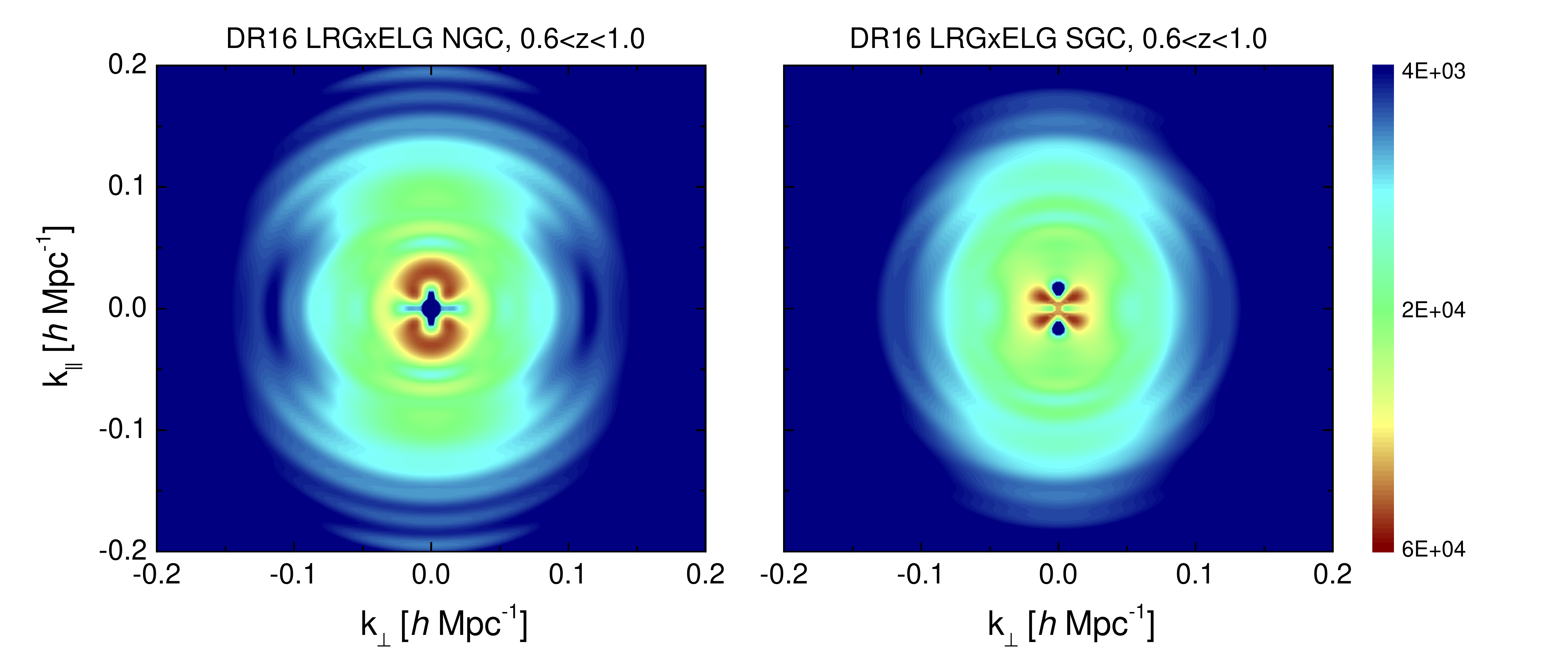}}
\caption{$P(k_{\bot},k_{||})$ of the cross power in the NGC (left) and SGC (right), reconstructed from the measured monopole, quadrupole and hexadecapole.}
\label{fig:Pk2D}
\end{figure*}

\begin{figure*}
\centering
{\includegraphics[scale=0.4]{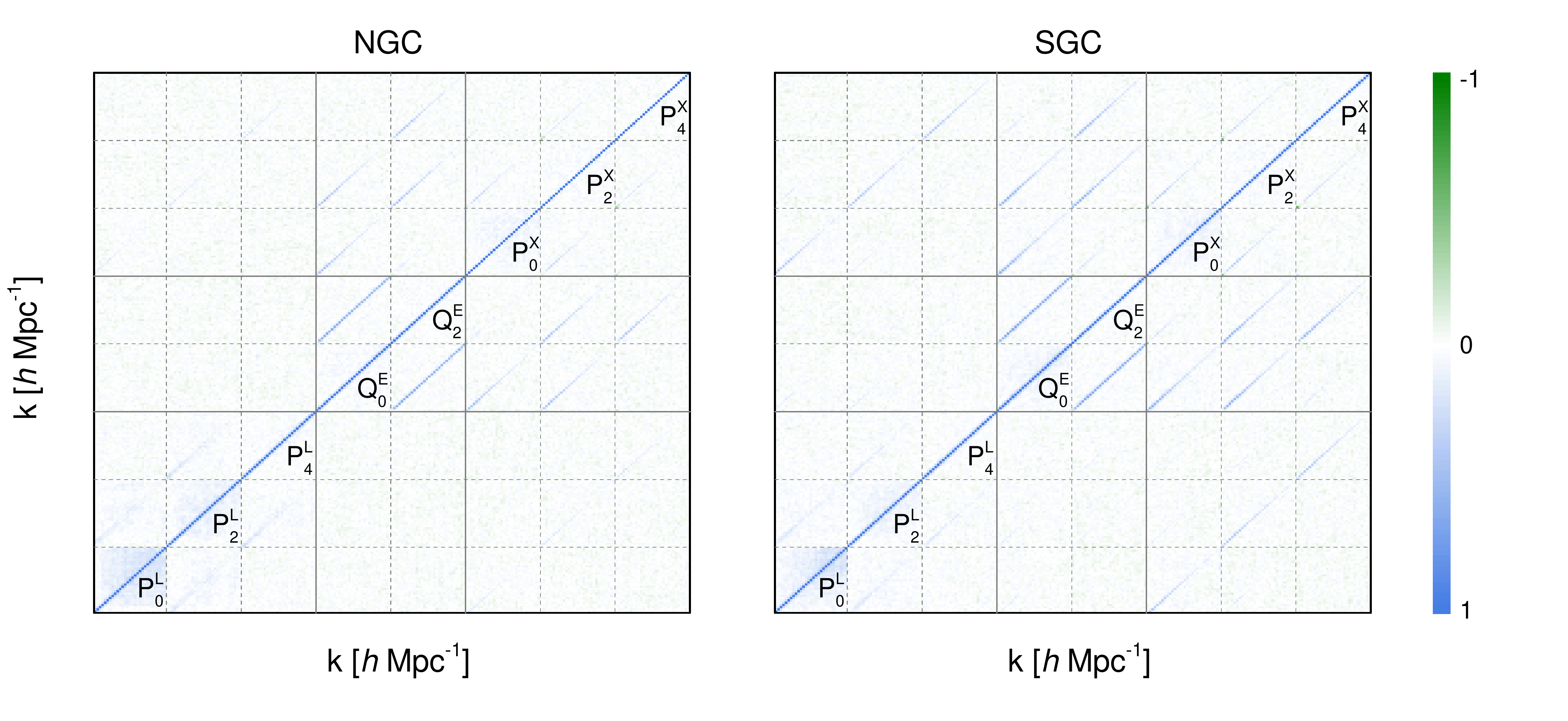}}
\caption{The correlation matrix for the power spectrum multipoles $P_{\ell}$ and $Q_{\ell}$ for the auto-power and cross-power spectrum of LRG and ELG, as illustrated in the legend. We omit the scales of the axes for brevity. For each sub-matrix (the minimal box in each panel), the $x$ and $y$ axes run from $0$ to $0.3\ \kunit$. The left and right panels are for the NGC and SGC, respectively.}
\label{fig:corr}
\end{figure*}

\begin{figure*}
\centering
{\includegraphics[scale=0.45]{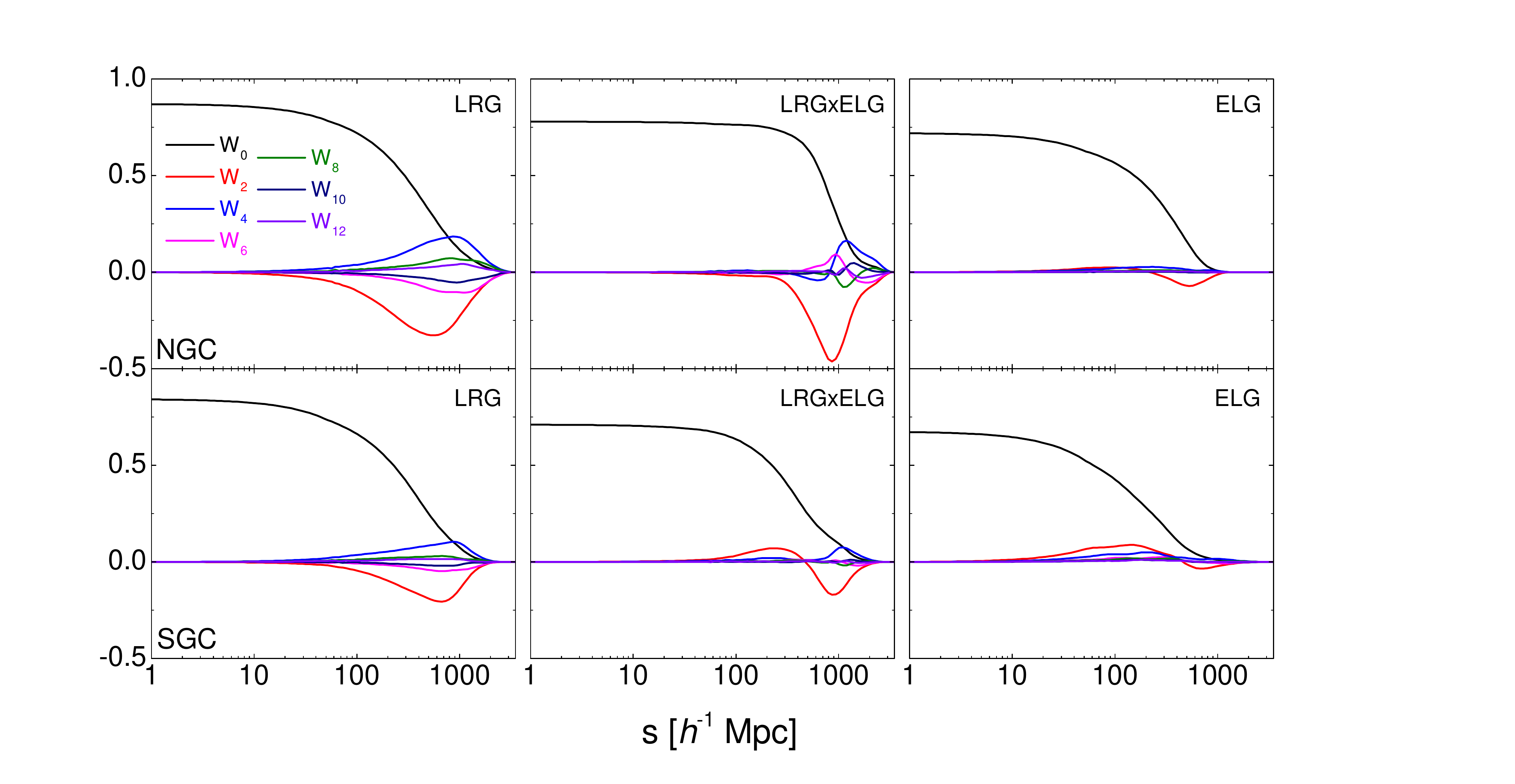}}
\caption{The window function multipoles (up to $\ell=12$) for the auto-power spectrum of LRG (left) and ELG (right), and for the cross power spectrum (middle). The upper and lower panels are for measurements from the NGC and SGC, respectively.}
\label{fig:window}
\end{figure*}

%% file: result.tex
\section{Results}
\label{sec:result}
This section is devoted to the main result of this work. We show our measurement of power spectrum multipoles from the EZmocks and from the DR16 galaxy sample, respectively, from which we derive a joint constraint on BAO and RSD parameters at multiple effective redshifts, after validating our pipeline using the EZmocks.

\begin{figure*}
\centering
{\includegraphics[scale=0.38]{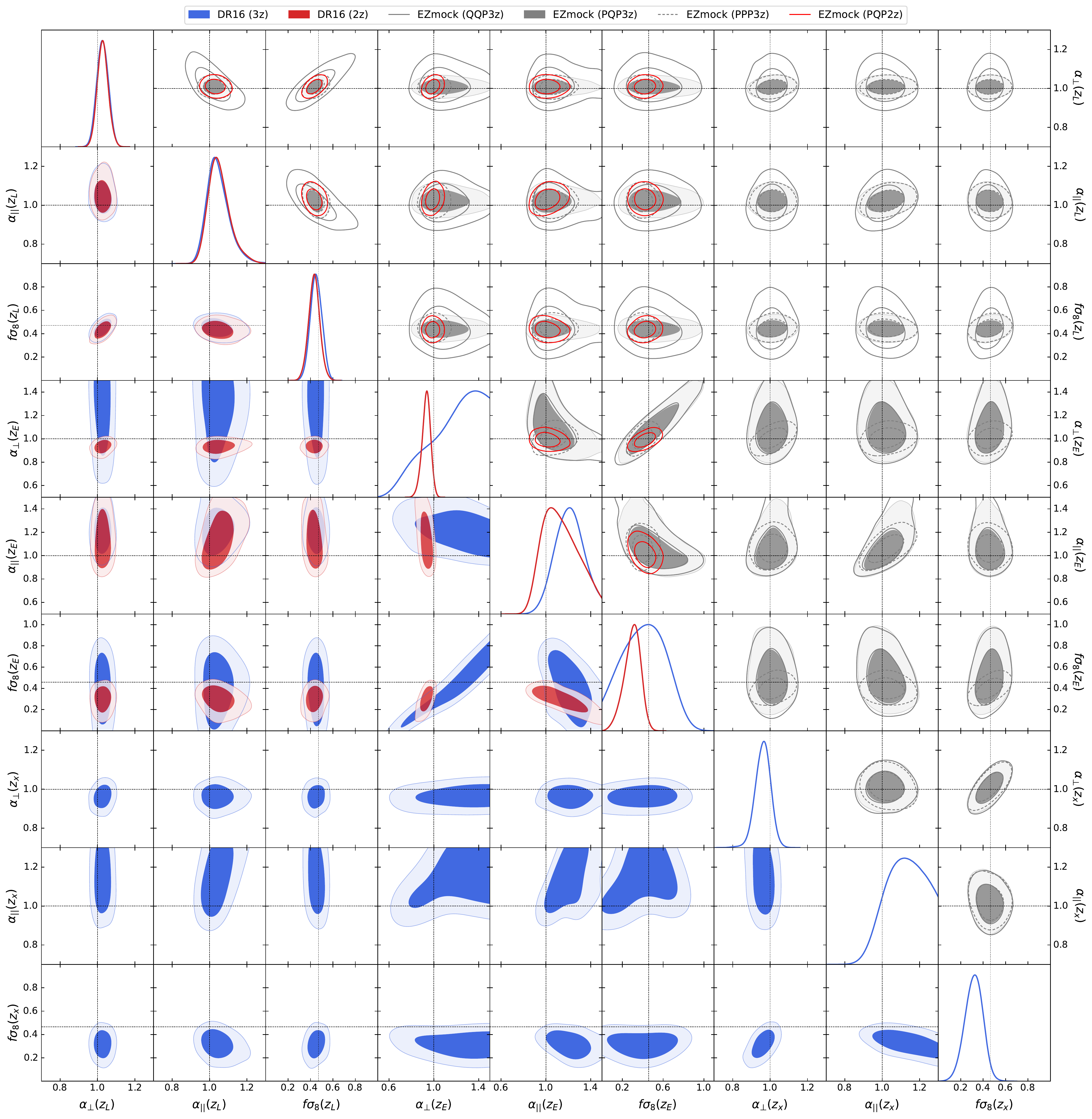}}
\caption{The one dimensional (1D) posterior distribution (panels on the diagonal) and 68 and 95\% CL contour plots for BAO and RSD parameters derived from the eBOSS DR16 (panels in the lower triangle) and EZmock catalogues (upper triangle), respectively. For measurements from the DR16 catalogue, results are derived at three (blue) and two (red) effective redshifts, respectively, denoted as DR16 ($3z$) and DR16 ($2z$) in the legend. \tc{For results derived from the mocks, measurements are preformed at three effective redshifts, from three different data combinations: QQP$3z$ (gray solid contours); PQP$3z$ (filled) and PPP$3z$ (gray dashed). We also preform the mock test at two effective redshifts, denoted as PQP$2z$ (red solid)}. The dashed horizontal and vertical lines illustrate the fiducial model used to produce the EZmocks, which is identical to that used for this work.}
\label{fig:contour}
\end{figure*}

\begin{table}
\caption{The mean and 68\% CL uncertainty of the BAO and RSD parameters measured from the DR16 LRG and ELG samples at three and two effective redshifts, respectively.}
\begin{center}

\begin{tabular}{ccc}

\hline\hline

Parameter   & Measurement ($3z$) & Measurement ($2z$)\\
\hline
$\abot(z=0.70)$ &  $1.028\pm0.031$ & $1.028\pm0.029$ \\
$\apar(z=0.70)$ & $1.047\pm0.063$  & $1.052\pm0.059$\\
$\fs(z=0.70)$ &  $0.450\pm0.051$  & $0.434\pm0.050$\\
$\bsILNz$ & $1.146\pm0.052$ & $1.186\pm0.058$ \\
$\bsILSz$ & $1.204\pm0.053$ & $1.233\pm0.044$\\
\hdashline
$\abot(z=0.77)$ & $0.961\pm0.041$ & $-$\\
$\apar(z=0.77)$ & $1.161^{+0.122}_{-0.159}$ & $-$ \\
$\fs(z=0.77)$ & $0.317\pm0.080$ & $-$ \\
\hdashline
$\abot(z=0.845)$ & $1.170^{+0.330}_{-0.091}$ & $0.933\pm0.038$ \\
$\apar(z=0.845)$ & $1.209\pm0.126$ & $1.130\pm0.155$\\
$\fs(z=0.845)$ &  $0.420\pm0.203$ & $0.297\pm0.081$\\
$\bsIENz$ & $0.867\pm0.098$ & $0.742\pm0.078$\\
$\bsIESz$ & $0.885\pm0.093$ & $0.767\pm0.070$\\
\hline
$\chi^2/{\rm DoF}$ & $205/(208-27)$ & $208/(208-25)$ \\
\hline\hline

\end{tabular}
\end{center}
\label{tab:DR16}
\end{table}%

\begin{figure}
\centering
{\includegraphics[scale=0.35]{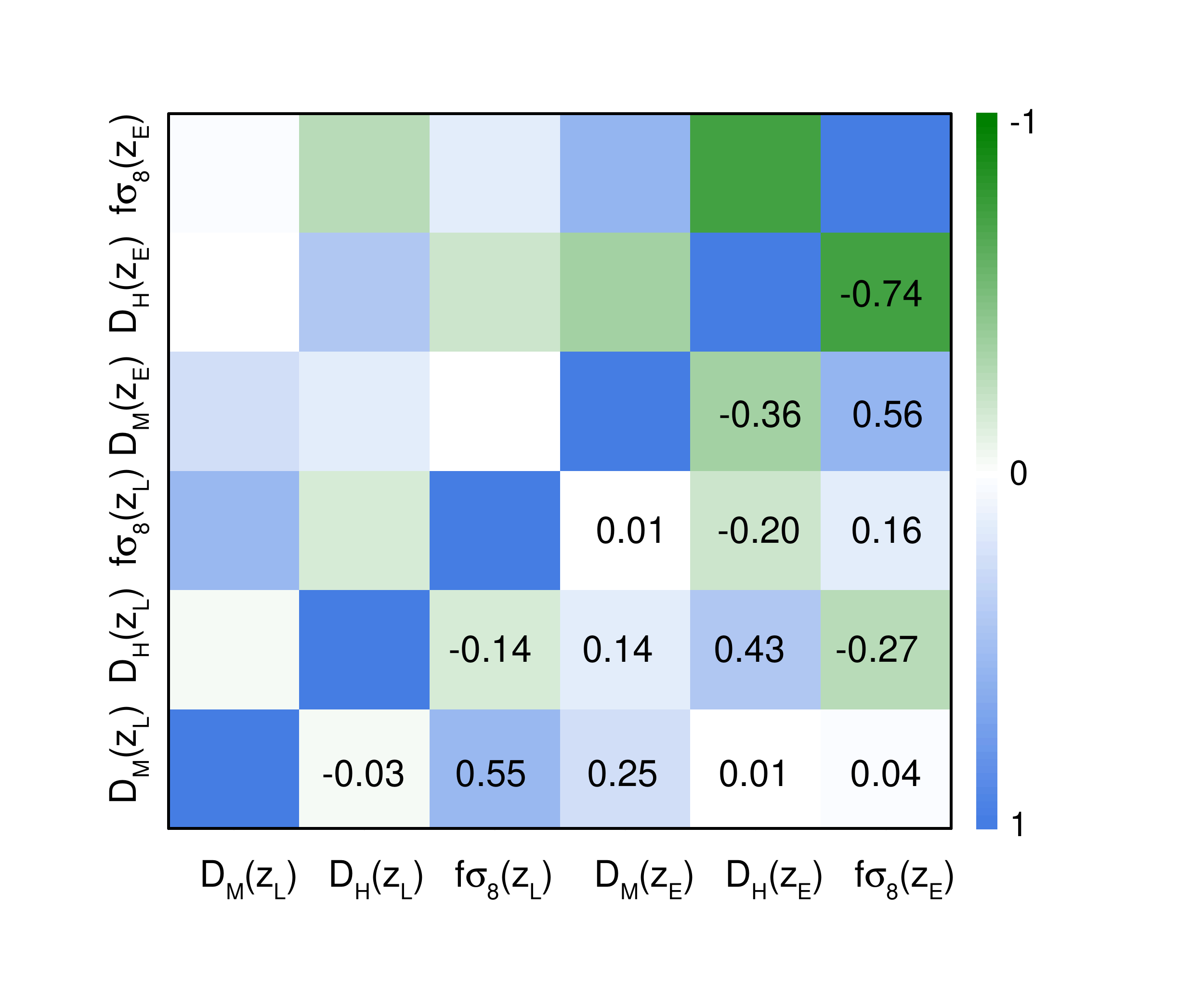}}
\caption{The correlation matrix for BAO and RSD parameters measured at two effective redshifts. The correlation coefficients (up to two digits) are marked in the figure for the ease of reading.}
\label{fig:corr_BAORSD}
\end{figure}

\begin{figure}
\centering
{\includegraphics[scale=0.25]{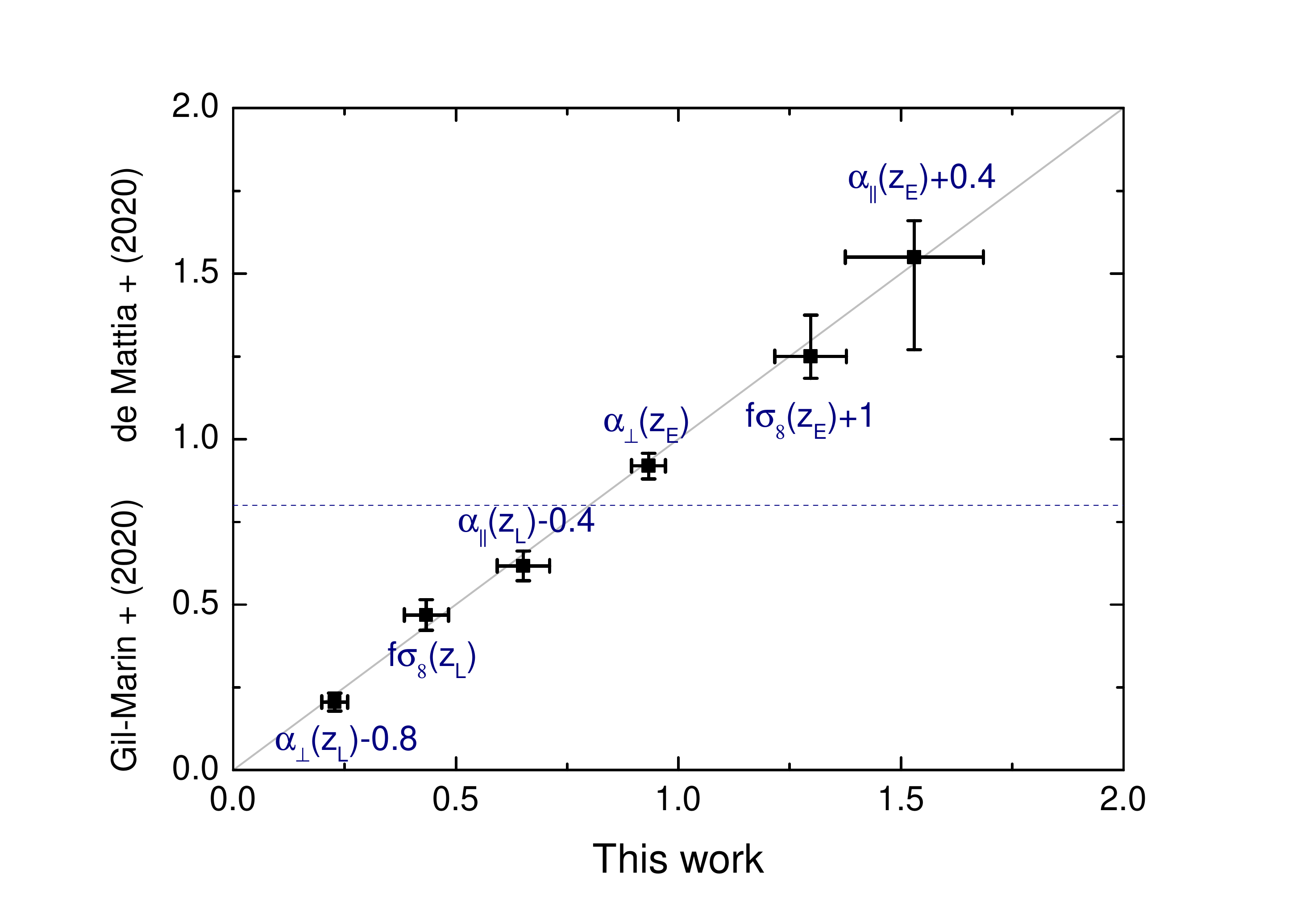}}
\caption{A comparison of BAO and RSD measurements between this work (x-axis) and the DR16 LRG analysis in $k$-space \citep{GilMarin2020} (first three data points on y-axis) and the DR16 ELG analysis in $k$-space \citep{deMattia2020} (last three data points). The data points are offset (as shown in the figure) for the ease of illustration. The diagonal dashed line shows $y=x$ for reference.}
\label{fig:compare}
\end{figure}

\begin{figure*}
\centering
{\includegraphics[scale=0.35]{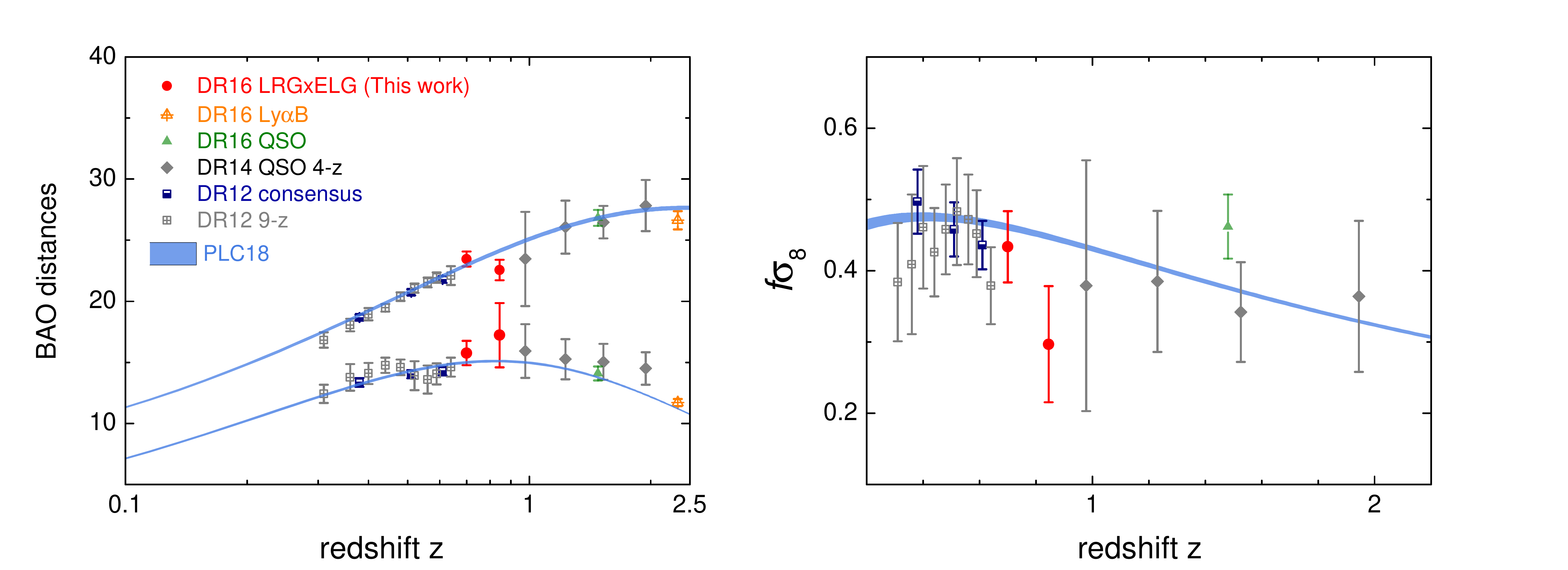}}
\caption{A compilation of BAO (left) and RSD measurements in recent years (right), including this work (filled red circle) and others derived from catalogues of DR16 Lyman-$\alpha$ forest (DR16 Ly$\alpha$B) \citep{duMas2020}, DR16 quasar (DR16 QSO) \citep{Hou2020,Neveux2020}, DR14 quasar (tomographic BAO and RSD measurements at $4$ effective redshifts) \citep{Zhao19}, BOSS DR12 (consensus BAO and RSD measurements at $3$ effective redshifts) \citep{alam}, DR12 (tomographic BAO and RSD measurements at nine effective redshifts) \citep{Zhaotomo16,Zheng19}. The shaded bands illustrate the 68\% CL constraint derived from Planck 2018 observations \citep{Planck18}, in the framework of a $\Lambda$CDM model. In the BAO figure, the upper and lower curves (and associated data points) are $\DM/\rd/\sqrt{z}+2$ and $\sqrt{z}\DH/\rd-2$, respectively.}
\label{fig:BAORSD}
\end{figure*}

\begin{figure*}
\centering
{\includegraphics[scale=0.38]{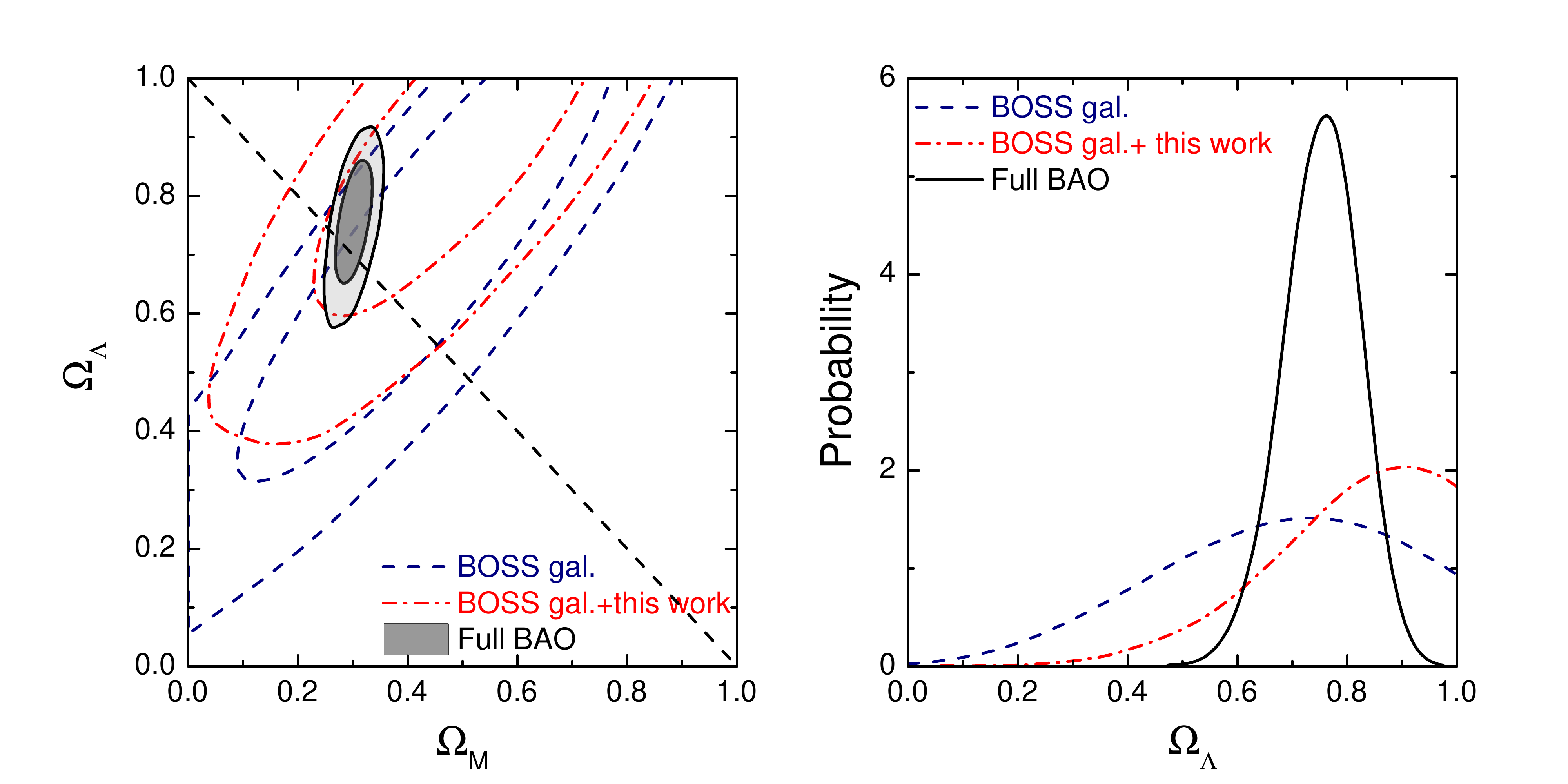}}
\caption{Constraints on $\Omega_{\rm M}$ and $\Omega_{\Lambda}$ using BAO observations alone. The left panel shows the $68$ and $95\%$ CL constraints derived from three datesets: (I) the BOSS DR12 BAO consensus result at three effective redshifts (`BOSS gal.'; blue dashed) \citep{alam}; (II) BOSS DR12 combined with this work (red dash-dotted; Note that we eliminated the BOSS BAO measurement for $z\in[0.6,0.8]$) in this combination, because BOSS galaxies in this redshift range are included in the DR16 LRG sample); (III) a further combination with BAO measurements using samples of MGS \citep{mgs}, 6dFGS \citep{BAO6dF}, eBOSS DR16 QSO \citep{Hou2020} and DR16 Lyman-$\alpha$ forest \citep{duMas2020} (`Full BAO'; gray filled). The right panel shows the 1D posterior distribution of $\Omega_{\Lambda}$ using three datasets. The posteriors are normalised so that the area under each curve is unity.}
\label{fig:Om_Ol}
\end{figure*}

\subsection{The power spectrum multipoles}

Figures \ref{fig:PQ_NGC} and \ref{fig:PQ_SGC} show the measurement of power spectrum multipoles $P_{\ell}$ and $\Q_{\ell}$ for the LRG and ELG samples in the NGC and SGC, respectively. The shaded bands illustrate the measurements (68\% CL uncertainty around the averaged power spectra) from $1000$ realisations of the EZmocks, and the data points with error bars are from the DR16 galaxy sample. Although measurements of the auto power spectrum in $P_{\ell}$ are presented and extensively discussed in \cite{GilMarin2020} and \cite{deMattia2020} for the LRG and ELG samples, respectively, they are included here for completeness, which is helpful for presenting and discussing the measurement of $Q_{\ell}$ and the cross power spectrum.

As expected, we see that $Q_{\ell}$ generally has larger uncertainties compared to $P_{\ell}$, because $P_{\ell+2}$, which is less well determined than $P_{\ell}$, is involved in $Q_{\ell}$. However, as claimed earlier, the unknown systematics in the data, if exists and couples to the $\mu=0$ mode, should be largely suppressed by using $Q_{\ell}$ instead. Interestingly, for $P_{\ell}$ measured from the ELG (NGC) sample, which is believed to be contaminated more by systematics than the SGC \citep{deMattia2020}, an offset between the DR16 sample and the EZmock in the quadrupole is clearly visible on scales at $k\lesssim0.06\kunit$, which might signal a component of unknown systematics. However, this glitch vanishes completely in the corresponding $Q_{\ell}$. 

The cross-power spectra in both galactic caps are successfully detected and well measured, although the signal to noise ratio is less than that of the auto-power spectra. We find that there is almost no qualitative difference in $P_{\ell}$ and $Q_{\ell}$ (up to $\ell=2$) for the cross power, reinforcing that the cross power is less affected by the systematics, as systematics for different tracers should be uncorrelated. Figure \ref{fig:Pk2D} shows the anisotropic cross power spectrum, which is reconstructed from the measured $P_0, P_2$ and $P_4$. A RSD pattern, which is the elongation of the clustering along $k_{||}$ is clearly visible in both galactic caps.

The correlation matrix for PQP is shown in Fig. \ref{fig:corr}, from which we see that $P^{\rm X}$ strongly correlates with $Q^{\rm E}$, but less with $Q^{\rm L}$. This is due to the fact that the ELG sample almost entirely overlaps with the LRG sample, so that the majority of the ELG contributes to the cross power. On the other hand, the LRG sample covers a much larger volume than the ELG, thus only a small fraction of the LRG is counted in the cross correlation. The correlation coefficient between $Q^{\rm L}$ and $Q^{\rm E}$ is relatively less (around $+0.3$), for the same reason.

Fig. \ref{fig:window} presents the window function multipoles measured from the random catalogues of the DR16 sample. As mentioned previously, the normalisation is performed in a way to match that for the power spectrum measurement, thus $W_0$ on small scales does not necessarily goes to unity. {\tc These window function multipoles are used to convolve the theoretical power spectrum prediction to account for the survey geometry following \citet{pkmask}, before a proper comparison between theory and data can be performed.}

\subsection{Demonstration using the EZmocks}

We perform a joint `$3z$' fit on the averaged power spectra of $1000$ realisations of the contaminated EZmocks using data vectors of PPP, PQP and QQP, respectively, for a validation and demonstration, and present the result in the upper triangle part of Fig. \ref{fig:contour}.

To start with, we notice that PPP and QQP provide the tightest and weakest constraint, respectively, and PQP is in between, as expected. The constraint from QQP and PQP are in excellent agreement with the expected values for all parameters, but the constraint from PPP can deviate by a noticeable amount, \eg, the constraints on $\apar(z_{\rm E})$ and $\fs(z_{\rm E})$ are higher or lower than the expected value by $\sim1\sigma$, due to the systematics in the ELG mock sample. This makes us decide not to use PPP for this work, although it provides the tighest constraint. QQP, on the other hand, unnecessarily trashes information of the LRG sample, which significantly dilutes the constraint at $z=0.70$. Due to these reasons, we choose to use PQP for presenting the primary result of this paper, as it is a reasonable compromise between retaining the constraining power of the data, and mitigating the systematics in the ELG sample. One point worth noting is that, the cross power, almost on its own, is able to provide a decent measurement at $\zeff=0.77$ with nearly no bias at all, which once again shows the robustness of the cross power against the systematics.

Mock tests with other data combinations and parametrisations are performed, and \eg, {\tc the case of PQP with `$2z$' is shown in red solid contours in the upper triangle. We also run tests with different cutoff scales for the power spectrum, and different widths of $k$ bins, and find that the choice adopted in this work is reasonably optimal. As illustrated,} our pipeline is well validated, \ie, the constraint derived in all cases are consistent with the expected ones well within the uncertainty, from all the tests. These mock tests also demonstrate that the cross power spectrum is informative, and more robust against systematics than the auto power spectrum.

\subsection{Measurements from the DR16 sample}

The `$3z$' and `$2z$' measurements from the DR16 galaxy sample (using PQP) are presented in Table \ref{tab:DR16} and in the lower triangle and the diagonal part of Fig. \ref{fig:contour}.

Measurement at $\zeff=0.70$ is well performed, thanks to the robust LRG observations. However, the `$3z$' measurement at $\zeff=0.77$ and $0.845$ are rather weak for some parameters, including all parameters for the ELG and $\apar(z_{\rm X})$, compared to those measured from the mean of mocks. This is largely because the ELG sample is subject to systematics {\tc including the} redshift failures, and unfortunately this kind of systematics affect both auto- and cross-correlations, so that the BAO feature gets distorted in the ELG auto- and cross correlation functions \citep{Tamone2020,Wang2020}. However, the cross power can constrain $\fs(z_{\rm X})$ fairly well, namely, $\fs(z_{\rm X})=0.317\pm0.080$, which is a $\sim4\sigma$ detection of the RSD signal, as visually illustrated in Fig. \ref{fig:Pk2D}. 

Due to the large correlation between $Q^{\rm E}$ and $P^{\rm X}$ as shown in Fig. \ref{fig:corr}, the BAO and RSD parameters measured at $\zX=0.77$ and $\zE=0.845$ are correlated. For example, ${\rm corr}[\apar(\zX),\apar(\zE)]=0.50, \ {\rm corr}[\apar(\zX),\abot(\zE)]=0.45$ and ${\rm corr}[\fs(\zX),\apar(\zE)]=-0.33$. This means that the weak constraints at $\zE$ can be improved, if the cross power spectrum is used to constrain parameters at $\zE$, which is designed as the `$2z$' measurement, as described in Sec. \ref{sec:parameter}.

Comparing the `$2z$' (top red layer in Fig. \ref{fig:contour}) with `$3z$' (bottom blue) measurements, we see that the constraint on all parameters at $\zE$ is significantly improved, primarily due to the contribution of the cross power spectrum. Specifically, $\abot(\zE)$, which is almost unconstrained in `$3z$' (it has no upper bound given the wide flat prior), is now measured at a precision of 4\% with a perfectly Gaussian distribution with the cross power combined in `$2z$'. The constraint on $\fs(\zE)$ is also improved by a significant amount, namely, the error bar is reduced by a factor of $2.5$. We notice that parameters at $\zL$ and $\zE$ are more correlated in the `$2z$' measurement, due to the cross power spectrum, as shown in Fig. \ref{fig:corr_BAORSD}.

Constraints on BAO and RSD parameters at $\zL$ and $\zE$ are extensively studied in companion papers of \cite{GilMarin2020} and \cite{deMattia2020}, respectively, using $P_{\ell}$ of the LRG and ELG samples separately. As an independent analysis using different methods in various aspects, we find that our results are fully consistent with these analyses within statistical uncertainties, as explicitly compared in Fig. \ref{fig:compare}. One noticeable difference, though, is seen for the uncertainty of parameters at $\zE$. The error bars derived in \cite{deMattia2020} are highly asymmetric, because of the non-Gaussian likelihood distribution. However, the posterior measured in this work is much closer to Gaussian, due to the contribution from the cross power. 

Final data product of this work is summarised in Eq. (\ref{eq:V}) and Fig. \ref{fig:BAORSD}, which are data vectors and covariance matrices for the BAO distances and $\fs$ at two redshifts,
\ba\label{eq:V} V(0.70) &=&  \l\{17.954\pm0.509,21.221\pm1.198,0.434\pm0.050 \r\} \nn \\ 
V(0.845) &=&  \l\{18.897\pm0.776,20.910\pm2.862,0.297\pm0.081 \r\} \nn \\
\ea where $ V \equiv \l\{\DM/\rd, \DH/\rd, \fs\r\}$.
These measurements are over-plotted with external measurements published in recent years, including one from the Planck2018 observations \citep{Planck18}, based on a $\Lambda$CDM model. Compared with the Planck result, our measurement at $\zE$ shows a {\tc roughly $2\sigma$ difference}, especially on $\DM/\rd$ and $\fs$. The same trend is independently found in \cite{deMattia2020} in the RSD measurement, which used a completely different scheme to mitigate the angular systematics. This may suggest interesting new physics beyond $\Lambda$CDM, although it may be subject to unknown residue systematics in the ELG sample, even after the mitigation by using the chained power spectrum and the cross power.

Projecting our BAO measurement onto the $\Omega_{\rm M}, \Omega_{\rm \Lambda}$ plane with $H_0\rd$ marginalised over as performed in \cite{Zhao19}, we find that the constraint is largely improved by combining our measurement with the that derived from the BOSS DR12 sample, namely, the error on $\Omega_{\rm \Lambda}$ is reduced by $22\%$, and the correlation with $\Omega_{\rm M}$ is lowered from $0.85$ to $0.75$, which raises the significance of $\Omega_{\rm \Lambda}>0$ from $2.95\sigma$ to $4.65\sigma$. Combining other BAO data to date, including the DR16 QSO and Lyman-$\alpha$ measurements, the nonzero $\Omega_{\rm \Lambda}$ is now favoured at a $\sim11\sigma$ confidence level, which is consistent with the multi-tracer analysis in the configuration space in a complementary paper \citep{Wang2020}.

\begin{table}
\caption{The constraints on $\Omega_{\rm M},\Omega_{\rm \Lambda}$ from BAO datasets, with $H_0\rd$ marginalised over.}
\begin{center}

\begin{tabular}{cccc}

\hline\hline
   & BOSS & BOSS + this work & Full BAO\\
\hline   
$\Omega_{\rm \Lambda}$ & $0.706\pm0.239$ & $0.864\pm0.186$ & $0.752\pm0.069$ \\
$\Omega_{\rm M}$ & $0.443\pm0.204$ & $0.480\pm0.172$ & $0.302\pm0.021$ \\
${\rm corr}[\Omega_{\rm M},\Omega_{\rm \Lambda}]$ & $0.85$ & $0.75$ & $0.55$ \\
S/N & $2.95$ & $4.65$ & $10.95$ \\
\hline\hline
\end{tabular}
\end{center}
\label{tab:FoM}
\end{table}%

\section*{Data Availability}
The data product of this work is publicly available at \url{https://github.com/icosmology/eBOSS_DR16_LRGxELG} and \url{https://www.sdss.org/science/final-bao-and-rsd-measurements/}

%% file: conclusion.tex
\section{Conclusion and Discussions}
\label{sec:conclusion}

eBOSS is a first galaxy survey to observe multiple tracers with a large overlap in the cosmic volume, which naturally motivated this work, as a study of BAO and RSD using multiple tracers in Fourier space.

This work is based on the eBOSS DR16 LRG and ELG samples in redshift range of $z\in[0.6,1.1]$, with more than $550,000$ galaxies in total. Being a first ELG sample for cosmological analysis in history, the DR16 ELG sample is analysed with particular care, to mitigate the systematics in the observations. For this purpose, we develop a new method using the chained power spectrum multipoles ($Q_{\ell}$), and has demonstrated using EZmocks that it can efficiently remove angular systematics. Being simply the algebraic difference between the normal power spectrum multipoles ($P_{\ell}$) with different orders, $Q_{\ell}$ is less well measured. Fortunately, the information loss in using $Q_{\ell}$ can be compensated by the cross power spectrum between the LRG and ELG samples, which itself is least affected by angular systematics.

We measure both $P_{\ell}$ and $Q_{\ell}$ for each tracer, as well as their cross power spectrum, and perform a joint BAO and RSD analysis at multiple redshifts after validating our pipeline using the EZmocks with systematics built in. {\tc Thanks to the quality of the eBOSS data, we are able to reach a $4\sigma$ detection of the cross power spectrum alone,} \ie, $\fs=0.317\pm0.080$, and find that adding cross-correlation in the analysis to the ELG sample can significantly boost the precision of the BAO and RSD measurement at $z=0.845$. Our final data product is summarised in Eq. (\ref{eq:BAORSDz}) and Fig. \ref{fig:corr_BAORSD}, which is a joint BAO and RSD measurement at $z=0.70$ and $z=0.845$, with the associated covariance matrix. Our measurement, combined with those measured from the eBOSS DR16 QSO \citep{Hou2020,Neveux2020} and Lyman-$\alpha$ sample \citep{duMas2020} and other galaxy catalogues at low redshits including the MGS \citep{mgs} and 6dFGS \citep{BAO6dF} samples, has raised the significance level of $\Omega_{\rm \Lambda}>0$ to $\sim11\sigma$. 

Methods developed in this work is directly applicable to forthcoming multi-tracer surveys including Dark Energy Spectroscopic Instrument (DESI) \citep{DESI}. {\tc Given the higher S/N of DESI, we expect the information loss to be reduced when using the chained power spectrum, with the cross power spectrum between different tracers included in the analysis. This makes it possible for mitigating angular systematics without degrading the statistical precision.}

% and in fact, measurement of the cross power spectrum from DESI with a high S/N can better reduce the information loss when using the chained power spectrum, which makes it possible for mitigating angular systematics without degrading the statistical precision. 

%% file: thanks.tex
\section*{Acknowledgements}

GBZ is supported by the National Key Basic Research and Development Program of China (No. 2018YFA0404503), and a grant of CAS Interdisciplinary Innovation Team. GBZ, YW and WBZ are supported by NSFC Grants 11925303, 11720101004, 11673025 and 11890691. YW is also supported by the Nebula Talents Program of NAOC {\tc and by the Youth Innovation Promotion Association CAS}. EMM has received funding from the European Research Council (ERC) under the European Unions Horizon 2020 research and innovation programme (grant agreement No 693024).

Funding for the Sloan Digital Sky Survey IV has been provided by the Alfred P. Sloan Foundation, the U.S. Department of Energy Office of Science, and the Participating Institutions. SDSS-IV acknowledges
support and resources from the Center for High-Performance Computing at
the University of Utah. The SDSS web site is \url{http://www.sdss.org/}. 

SDSS-IV is managed by the Astrophysical Research Consortium for the 
Participating Institutions of the SDSS Collaboration including the 
Brazilian Participation Group, the Carnegie Institution for Science, 
Carnegie Mellon University, the Chilean Participation Group, the French Participation Group, Harvard-Smithsonian Center for Astrophysics, 
Instituto de Astrof\'isica de Canarias, The Johns Hopkins University, Kavli Institute for the Physics and Mathematics of the Universe (IPMU) / 
University of Tokyo, the Korean Participation Group, Lawrence Berkeley National Laboratory, 
Leibniz Institut f\"ur Astrophysik Potsdam (AIP),  
Max-Planck-Institut f\"ur Astronomie (MPIA Heidelberg), 
Max-Planck-Institut f\"ur Astrophysik (MPA Garching), 
Max-Planck-Institut f\"ur Extraterrestrische Physik (MPE), 
National Astronomical Observatories of China, New Mexico State University, 
New York University, University of Notre Dame, 
Observat\'ario Nacional / MCTI, The Ohio State University, 
Pennsylvania State University, Shanghai Astronomical Observatory, 
United Kingdom Participation Group,
Universidad Nacional Aut\'onoma de M\'exico, University of Arizona, 
University of Colorado Boulder, University of Oxford, University of Portsmouth, 
University of Utah, University of Virginia, University of Washington, University of Wisconsin, 
Vanderbilt University, and Yale University.

This work made use of three supercomputing facilities, including I) The Laohu cluster supported by National Astronomical Observatories, Chinese Academy of Sciences; II) The National Energy Research Scientific Computing Center, a DOE Office of Science User Facility supported by the Office of Science of the U.S. Department of Energy under Contract No. DE-AC02-05CH11231; and III) The UK Sciama High Performance Computing cluster supported by the ICG, SEPNet and the University of Portsmouth. We used the {\tt FFTW}\footnote{\url{http://www.fftw.org}} library and the {\tt jobfork}\footnote{\url{https://github.com/cheng-zhao/jobfork}} tool for numerical calculation.

%% file: appendix.tex
\onecolumn
\appendix
\section{The extended TNS model for the cross power spectrum}
\label{sec:xTNS}
\subsection{Preliminaries}
\label{sec:preliminaries}
%%%%%%%%%%%%%%%%%%%%%%%%%%%%%%%%%%%%%%%%%%%%%%%%%%%%%%%%%%%%%%%%%%%%%%%
%%%%%%%%%%%%%%%%%%%%%%%%%%%%%%%%%%%%%%%%%%%%%%%%%%%%%%%%%%%%%%%%%%%%%%%

Throughout the report, we work with the distant-observer limit, and assume that the line-of-sight direction is parallel to the $z$-axis. Then the {\it observed} redshift space may be written as
%%%%%%%%%%%%%%%%%%%%%%%%%%%%%%%%%%%%%%%%%%%%%%%%%%%
\begin{align}
\bfs=\bfr - f\,u_z(\bfr)\hat{z},
\end{align}
%%%%%%%%%%%%%%%%%%%%%%%%%%%%%%%%%%%%%%%%%%%%%%%%%%%
where the quantity $u_z$ is the normalised velocity field along the line-of-sight, defined by $u_z\equiv-v_z/(aH\,f)$. The density field in observed redshift space, $\deltas$, is expressed in Fourier space as
%%%%%%%%%%%%%%%%%%%%%%%%%%%%%%%%%%%%%%%%%%%%%%%%%%%
\begin{align}
 \deltas(\bfk) = \int d^3\bfr \Bigl\{\delta(\bfr)+f\,\nabla_z u_z(\bfr)\Bigr\}\,e^{i\,\{ \bfk\cdot\bfr- k\mu\,f\, u_z\}}
\end{align}
%%%%%%%%%%%%%%%%%%%%%%%%%%%%%%%%%%%%%%%%%%%%%%%%%%%
with $\mu\equiv k_z/k$.

We are particularly interested in the cross correlation between the different samples (with different bias parameter). We denote the number density fluctuation of the objects $A$ and $B$ by $\deltaA$ and $\deltaB$. {\tc Also, we consider that the velocity for each object do not simply trace the underlying mass density field, \ie, we generically allow for velocity biases for each tracer, and is labeled as $u_{\rm A,B}$}. Then, the cross power spectrum is expressed as 
%%%%%%%%%%%%%%%%%%%%%%%%%%%%%%%%%%%%%%%%%%%%%%%%%%%
\begin{align}
 P^{\rm(S)}(\bfk)=\int d^3\bfx\,e^{i\bfk\cdot\bfx}\,\Bigl\langle
e^{-i\,k\mu\,(f\Delta u_z+\Delta \epsilon)}\Bigl[\deltaA(\bfr)+f\nabla_z u_{{\rm A},z}(\bfr)\Bigr]\Bigl[\deltaB(\bfr')+f\nabla_z u_{{\rm B},z}(\bfr')\Bigr]
\Bigr\rangle
\label{eq:pkred_exact}
\end{align}
%%%%%%%%%%%%%%%%%%%%%%%%%%%%%%%%%%%%%%%%%%%%%%%%%%%
with $\bfx=\bfr-\bfr'$. We here define  
%%%%%%%%%%%%%%%%%%%%%%%%%%%%%%%%%%%%%%%%%%%%%%%%%%%
\begin{align}
 \Delta u_z\equiv u_{{\rm A},z}(\bfr)-u_{{\rm B},z}(\bfr').
\end{align}
%%%%%%%%%%%%%%%%%%%%%%%%%%%%%%%%%%%%%%%%%%%%%%%%%%%

%%%%%%%%%%%%%%%%%%%%%%%%%%%%%%%%%%%%%%%%%%%%%%%%%%%%%%%%%%%%%%%%%%%%%%%
%%%%%%%%%%%%%%%%%%%%%%%%%%%%%%%%%%%%%%%%%%%%%%%%%%%%%%%%%%%%%%%%%%%%%%%
\subsection{Modeling redshift-space cross power spectrum at weakly nonlinear regime}
\label{sec:modeling}
%%%%%%%%%%%%%%%%%%%%%%%%%%%%%%%%%%%%%%%%%%%%%%%%%%%%%%%%%%%%%%%%%%%%%%%
%%%%%%%%%%%%%%%%%%%%%%%%%%%%%%%%%%%%%%%%%%%%%%%%%%%%%%%%%%%%%%%%%%%%%%%

To derive the expression relevant in the weakly nonlinear regime, we follow Ref.~\cite{TNS}, and rewrite Eq.~(\ref{eq:pkred_exact}) with 
%%%%%%%%%%%%%%%%%%%%%%%%%%%%%%%%%%%%%%%%%%%%%%%%%%%
\begin{align}
 P^{\rm(S)}(\bfk)=\int d^3\bfx\,e^{i\bfk\cdot\bfx}
\Bigl\langle
e^{j_1\,A_1}\,A_2\,A_3
\Bigr\rangle
\end{align}
%%%%%%%%%%%%%%%%%%%%%%%%%%%%%%%%%%%%%%%%%%%%%%%%%%%
with the quantities $j_1$, $A_i$ given by
%%%%%%%%%%%%%%%%%%%%%%%%%%%%%%%%%%%%%%%%%%%%%%%%%%%
\begin{align}
& j_1 = -i\,k\mu,
\nonumber
\\
 & A_1=f\,\Delta u_z
\nonumber
\\
 & A_2=\deltaA(\bfr) +f\nabla_z\,u_{{\rm A},z}(\bfr),
\nonumber
\\
 & A_3=\deltaB(\bfr') +f\nabla_z\,u_{{\rm B},z}(\bfr').
\nonumber
\end{align}
%%%%%%%%%%%%%%%%%%%%%%%%%%%%%%%%%%%%%%%%%%%%%%%%%%%
Then, with a help of cumulant expansion theorem, we obtain
%%%%%%%%%%%%%%%%%%%%%%%%%%%%%%%%%%%%%%%%%%%%%%%%%%%
\begin{align}
 P^{\rm(S)}(\bfk)=\int d^3\bfx\,e^{i\bfk\cdot\bfx}
\exp\Bigl\{\langle e^{j_1A_1}\rangle_c\Bigr\}\,\Bigl[
\bigl\langle e^{j_1A_1}A_2A_3\bigr\rangle_c+
\bigl\langle e^{j_1A_1}A_2\bigr\rangle_c\bigl\langle e^{j_1A_1}A_3\bigr\rangle_c
\Bigr].
\label{eq:pkred_exact2}
\end{align}
%%%%%%%%%%%%%%%%%%%%%%%%%%%%%%%%%%%%%%%%%%%%%%%%%%%
Here, $\langle\cdots\rangle_c$ indicates the cumulant.

As it is clear from the expression, the exponential prefactor 
$\exp\bigl\{\langle e^{j_1A_1}\rangle_c\bigr\}$ can be non-perturbative, and it leads to a strong damping even at large scales. We thus keep it untouched. But, at weakly nonlinear scales, we may expand the rest of the terms regarding $j_1$ as a small expansion parameter. Up to the order of $\mathcal{O}(j_1^2)$, we obtain
%%%%%%%%%%%%%%%%%%%%%%%%%%%%%%%%%%%%%%%%%%%%%%%%%%%
\begin{align}
 P^{\rm(S)}(\bfk)\simeq\int d^3\bfx\,e^{i\bfk\cdot\bfx}
\exp\Bigl\{\langle e^{j_1A_1}\rangle_c\Bigr\}\,\Bigl[
\bigl\langle A_2A_3\bigr\rangle_c+j_1\bigl\langle A_1A_2A_3\bigr\rangle_c+
j_1^2 \bigl\langle A_1A_2\bigr\rangle_c\bigl\langle A_1A_3\bigr\rangle_c+\cdots
\Bigr].
\end{align}
%%%%%%%%%%%%%%%%%%%%%%%%%%%%%%%%%%%%%%%%%%%%%%%%%%%
Here, the term $\frac{1}{2}j_1^2\langle A_1^2A_2A_3\rangle_c$ is ignored according to \cite{TNS}. For more simplification, we shall assume that $\exp\bigl\{\langle e^{j_1A_1}\rangle_c\bigr\}$ is independent of separation $x$, and is expressed as (even) function of $k\mu$. With this assumption/ansatz, the model of redshift-space cross power spectrum, $P^{\rm(S)}_{\rm AB}$, is given by
%%%%%%%%%%%%%%%%%%%%%%%%%%%%%%%%%%%%%%%%%%%%%%%%%%%
\begin{align}
 P^{\rm(S)}_{\rm AB}(\bfk)=D_{\rm FoG}(k\mu\sigmav)\,\Bigl[\Pkaiser(\bfk) + \Aterm(\bfk) + \Bterm(\bfk)\Bigr]
\label{eq:pkred_model}
\end{align}
%%%%%%%%%%%%%%%%%%%%%%%%%%%%%%%%%%%%%%%%%%%%%%%%%%%
with
%%%%%%%%%%%%%%%%%%%%%%%%%%%%%%%%%%%%%%%%%%%%%%%%%%%
\begin{align}
\Pkaiser(\bfk) &=\int d^3\bfx\,e^{i\bfk\cdot\bfx} \bigl\langle A_2A_3\bigr\rangle_c, 
\nonumber
\\
\Aterm(\bfk) & = j_1\int d^3\bfx\,e^{i\bfk\cdot\bfx} \bigl\langle A_1A_2A_3\bigr\rangle_c,
\nonumber
\\
\Bterm(\bfk) & = j_1^2\int d^3\bfx\,e^{i\bfk\cdot\bfx} \bigl\langle A_1A_2\Bigr\rangle\Bigl\langle A_1A_3\bigr\rangle_c.
\end{align}
%%%%%%%%%%%%%%%%%%%%%%%%%%%%%%%%%%%%%%%%%%%%%%%%%%%
Below, we explicitly write down the expression of each term in the bracket. In what follows, we assume the linear bias for $\deltaA$ and $\deltaB$, and rewrite them with $\bA\,\delta$ and $\bB\,\delta$, respectively. Similarly, assuming the linear relation, we may write biased velocity field as $\bfu_{\rm A,B}=c_{\rm A,B} \,\bfu$. With the velocity-divergence field $\theta$ defined by $\theta=\nabla\cdot \bfu=-\nabla\cdot\bfv/(afH)$, we then have:
%%%%%%%%%%%%%%%%%%%%%%%%%%%%%%%%%%%%%%%%%%%%%%%%%%%
\begin{align}
\Pkaiser(k,\mu) & = \bA \bB\,P_{\delta\delta}(k)+f\,\mu^2(\bA \cB +\bB \cA) P_{\delta\theta}(k)+f^2\,\mu^4\,\cA \cB\,P_{\theta\theta}(k),
\label{eq:generalized_Kaiser}
%\end{align}
%%%%%%%%%%%%%%%%%%%%%%%%%%%%%%%%%%%%%%%%%%%%%%%%%%%
\\
%%%%%%%%%%%%%%%%%%%%%%%%%%%%%%%%%%%%%%%%%%%%%%%%%%%
%\begin{align}
\Aterm(k,\mu) &= 
k\mu \,f  \int\frac{d^3\bfp}{(2\pi)^3}\frac{p_z}{p^2}
\Bigl\{\cA\,\bispecAterm(\bfp,\bfk-\bfp,-\bfk)-
\cB\, \bispecAterm(\bfp,\bfk,-\bfk-\bfp)\Bigr\},
\label{eq:A_term}
%\end{align}
%%%%%%%%%%%%%%%%%%%%%%%%%%%%%%%%%%%%%%%%%%%%%%%%%%%
\\
%%%%%%%%%%%%%%%%%%%%%%%%%%%%%%%%%%%%%%%%%%%%%%%%%%%
%\begin{align}
 \Bterm(k,\mu) &= (k\mu\,f)^2 \cA\cB\,\int\frac{d^3\bfp\,d^3\bfq}{(2\pi)^3} 
\delta_{\rm D}(\bfk-\bfp-\bfq)\,\tilde{F}_{\rm A}(\bfp)\tilde{F}_{\rm B}(\bfq),
\label{eq:B_term}
\end{align}
%%%%%%%%%%%%%%%%%%%%%%%%%%%%%%%%%%%%%%%%%%%%%%%%%%%
where the quantities $\bispecAterm$, $\tilde{F}_{\rm X}$ ($X=$A or B) are the cross bispectrum and power spectrum, respectively, defined by
%%%%%%%%%%%%%%%%%%%%%%%%%%%%%%%%%%%%%%%%%%%%%%%%%%%
\begin{align}
&(2\pi)^3\,\delta_{\rm D}(\bfk_1+\bfk_2+\bfk_3)\bispecAterm(\bfk_1,\bfk_2,\bfk_3)
\nonumber
\\
& \qquad =\Bigl\langle
\theta(\bfk_1)\bigl\{\bA\,\delta(\bfk_2)+\cA\,f\Bigl(\frac{k_{2,z}}{k_2}\Bigr)^2\theta(\bfk_2)\bigr\}\bigl\{\bB\,\delta(\bfk_3)+\cB\,f\Bigl(\frac{k_{3,z}}{k_3}\Bigr)^2\theta(\bfk_3)\bigr\}\Bigr\rangle.
\label{eq:bispec_Aterm}
%\end{align}
%%%%%%%%%%%%%%%%%%%%%%%%%%%%%%%%%%%%%%%%%%%%%%%%%%%
\\
%%%%%%%%%%%%%%%%%%%%%%%%%%%%%%%%%%%%%%%%%%%%%%%%%%%
%\begin{align}
& \tilde{F}_{\rm X}(\bfp)=\frac{p_z}{p^2}\,\Bigl\{b_{\rm X} \,P_{\delta\theta}(p)+c_{\rm X}\,f\,\Bigl(\frac{p_z^2}{p^2}\Bigr)^2\,P_{\theta\theta}(p)
\Bigr\}.
\label{eq:pk_F_Aterm}
\end{align}
%%%%%%%%%%%%%%%%%%%%%%%%%%%%%%%%%%%%%%%%%%%%%%%%%%%

We will derive below the explicit expressions for $\Aterm$ and $\Bterm$, which are given in powers of $\mu$ and $f$.

%%%%%%%%%%%%%%%%%%%%%%%%%%%%%%%%%%%%%%%%%%%%%%%%%%%%%%%%%%%%%%%%%%%%%%%
%%%%%%%%%%%%%%%%%%%%%%%%%%%%%%%%%%%%%%%%%%%%%%%%%%%%%%%%%%%%%%%%%%%%%%%
\subsubsection{$\Aterm$ term}
\label{subsec:A_term}
%%%%%%%%%%%%%%%%%%%%%%%%%%%%%%%%%%%%%%%%%%%%%%%%%%%%%%%%%%%%%%%%%%%%%%%
%%%%%%%%%%%%%%%%%%%%%%%%%%%%%%%%%%%%%%%%%%%%%%%%%%%%%%%%%%%%%%%%%%%%%%%

The bispectrum $\bispecAterm$ given at Eq.~(\ref{eq:bispec_Aterm}) is related to the real-space matter bispectra, $B_{abc}$,  defined by $\langle\Phi_a(\bfk_1)\Phi_b(\bfk_2)\Phi_c(\bfk_3)\rangle=(2\pi)^3\delta_{\rm D}(\bfk_1+\bfk_2+\bfk_3)\,B_{abc}(\bfk_1,\bfk_2,\bfk_3)$ with doublet $\Phi_a=(\delta,\,\theta)$.  It is given by
%%%%%%%%%%%%%%%%%%%%%%%%%%%%%%%%%%%%%%%%%%%%%%%%%%%%%%%%%%%%%%%%%%%%%%%
\begin{align}
 \bispecAterm(\bfk_1,\bfk_2,\bfk_3)&=
\textcolor{black}{\bA\bB\,B_{211}(\bfk_1,\bfk_2,\bfk_3)+
\cA\cB\,f^2\Bigl(\frac{k_{2,z}}{k_2}\Bigr)^2\Bigl(\frac{k_{3,z}}{k_3}\Bigr)^2\,B_{222}(\bfk_1,\bfk_2,\bfk_3)}
\nonumber
\\
&+\textcolor{black}{\bA\cB\,f\Bigl(\frac{k_{3,z}}{k_3}\Bigr)^2\,B_{212}(\bfk_1,\bfk_2,\bfk_3)+
\bB\cA\,f\Bigl(\frac{k_{2,z}}{k_2}\Bigr)^2\,B_{221}(\bfk_1,\bfk_2,\bfk_3)}
\nonumber
\\
&\equiv
\textcolor{black}{\bispecAterm^{\rm(sym)}(\bfk_1,\bfk_2,\bfk_3)} + 
\textcolor{black}{\bispecAterm^{\rm(non\mbox{-}sym)}(\bfk_1,\bfk_2,\bfk_3)}
\end{align}
%%%%%%%%%%%%%%%%%%%%%%%%%%%%%%%%%%%%%%%%%%%%%%%%%%%%%%%%%%%%%%%%%%%%%%%
Note that the first line at RHS or $\bispecAterm^{\rm(sym)}$ is symmetric under $\bfk_2\leftrightarrow\bfk_3$, but the second line or $\bispecAterm^{\rm(non\mbox{-}sym)}$ is not, and can become symmetric only in the auto-power spectrum (i.e., $\bA=\bB$ and $\cA=\cB$). This asymmetry gives rise to non-trial contribution, which makes the $\Aterm$ term different from that in the auto-power spectrum case. 

To derive the explicit expressions of the $\Aterm$ term in powers of $\mu$ and $f$, we rewrite Eq.~(\ref{eq:A_term}) as 
%%%%%%%%%%%%%%%%%%%%%%%%%%%%%%%%%%%%%%%%%%%%%%%%%%%%%%%%%%%%%%%%%%%%%%%
\begin{align}
\Aterm(k,\mu) &= 
 k\mu \,f  \int\frac{d^3\bfp}{(2\pi)^3}
\Bigl\{\cA\,\frac{p_z}{p^2}\,\bispecAterm^{\rm(sym)}(\bfp,\bfk-\bfp,-\bfk)+
\cB\, \frac{k_z-p_z}{|\bfk-\bfp|^2}\,\bispecAterm^{\rm(sym)}(\bfk-\bfp,\bfp,-\bfk)\Bigr\}
\nonumber
\\
&+ k\mu \,f  \int\frac{d^3\bfp}{(2\pi)^3}
\Bigl\{\cA\,\frac{p_z}{p^2}\,\bispecAterm^{\rm(non\mbox{-}sym)}(\bfp,\bfk-\bfp,-\bfk)+
\cB\, \frac{k_z-p_z}{|\bfk-\bfp|^2}\,\bispecAterm^{\rm(non\mbox{-}sym)}(\bfk-\bfp,-\bfk,\bfp)\Bigr\}
\end{align}
%%%%%%%%%%%%%%%%%%%%%%%%%%%%%%%%%%%%%%%%%%%%%%%%%%%%%%%%%%%%%%%%%%%%%%%
where $\bispecAterm^{\rm(sym)}$ and $\bispecAterm^{\rm(non\mbox{-}sym)}$ are defined below:
%%%%%%%%%%%%%%%%%%%%%%%%%%%%%%%%%%%%%%%%%%%%%%%%%%%%%%%%%%%%%%%%%%%%%%%
\begin{align}
 \bispecAterm^{\rm(sym)}(\bfk_1,\bfk_2,\bfk_3)&=\bA\bB\,B_{211}(\bfk_1,\bfk_2,\bfk_3)+
\cA\cB\,f^2\Bigl(\frac{k_{2,z}}{k_2}\Bigr)^2\Bigl(\frac{k_{3,z}}{k_3}\Bigr)^2\,B_{222}(\bfk_1,\bfk_2,\bfk_3),
\nonumber
\\
 \bispecAterm^{\rm(non\mbox{-}sym)}(\bfk_1,\bfk_2,\bfk_3)&=\bA\cB\,f\Bigl(\frac{k_{3,z}}{k_3}\Bigr)^2\,B_{212}(\bfk_1,\bfk_2,\bfk_3)+\bB\cA\,f\Bigl(\frac{k_{2,z}}{k_2}\Bigr)^2\,B_{221}(\bfk_1,\bfk_2,\bfk_3).
\end{align}
%%%%%%%%%%%%%%%%%%%%%%%%%%%%%%%%%%%%%%%%%%%%%%%%%%%%%%%%%%%%%%%%%%%%%%%
With the form given above, the $\Aterm$ is expanded as
%%%%%%%%%%%%%%%%%%%%%%%%%%%%%%%%%%%%%%%%%%%%%%%%%%%%%%%%%%%%%%%%%%%%%%%
\begin{align}
& \Aterm(k,\mu)=\frac{k^3}{(2\pi)^2}\sum_{n=1}^3\sum_{a,b}^2\mu^{2n}\,f^{a+b-1}\,
\int_0^\infty dr\int_{-1}^1 dx\,
\nonumber
\\
&\qquad\times
\Bigl\{
A^n_{ab}(r,x)\,B_{2ab}(\bfp,\bfk-\bfp,-\bfk) +
\tilde{A}^n_{ab}(r,x)\,B_{2ab}(\bfk-\bfp,\bfp,-\bfk) +
\hat{A}^n_{ab}(r,x)\,B_{2ab}(\bfk-\bfp,-\bfk,\bfp)
\Bigr\}, 
\end{align}
%%%%%%%%%%%%%%%%%%%%%%%%%%%%%%%%%%%%%%%%%%%%%%%%%%%%%%%%%%%%%%%%%%%%%%%
where we define $r=p/k$ and $x=(\bfk\cdot\bfp )/(k p)$. 
Then, according to Appendix B of \cite{TNS}, the coefficients $A^n_{ab}$, $\tilde{A}^a_{ab}$, and $\hat{A}^a_{ab}$ are derived, and the non-vanishing coefficients are expressed as follows:
%%%%%%%%%%%%%%%%%%%%%%%%%%%%%%%%%%%%%%%%%%%%%%%%%%%%%%%%%%%%%%%%%%%%%%%
\begin{align}
& A^1_{11}=r\, x\, \bA \bB \cA,\quad
 A^1_{21}=-\frac{r^2(-2+3rx)(x^2-1)}{2(1+r^2-2rx)}\,\bB \cA^2,\quad
 A^2_{12}=r\, x\, \bA \cA \cB, \quad
\nonumber
\\
&  A^2_{21}=\frac{r(2 x+r (2-6 x^2)+r^2 x (-3+5 x^2))}{2(1+r^2-2rx)}\, \bB \cA^2, \quad
A^2_{22}=-\frac{r^2 (-2+3 r x) (x^2-1)}{2(1+r^2-2rx)}\,  \cA^2 \cB, \quad
\nonumber
\\
& A^3_{22}=\frac{r (2 x+r (2-6 x^2+r x (-3+5 x^2))) }{2(1+r^2-2rx)}\,  \cA^2 \cB\quad
\nonumber
\\
& \tilde{A}^1_{11}=-\frac{r^2(-1+rx)}{(1+r^2-2rx)}\,\bA \bB \cB,\quad
 \tilde{A}^2_{22}=\frac{ r^2 (-1+3 r x) (x^2-1)}{2(1+r^2-2rx)}\, \cA \cB^2,\quad
 \tilde{A}^3_{22}=\frac{r^2 (-1+3 r x+3 x^2-5 r x^3) }{2(1+r^2-2rx)}\, \cA \cB^2 ,\quad
\nonumber
\\
& \hat{A}^1_{12}=\frac{r^2(-1+3rx)(x^2-1)}{2(1+r^2-2rx)}\,\bA \cB^2,\quad
 \hat{A}^2_{12}=-\frac{r^2 (1-3 x^2+r x (-3+5 x^2))}{2(1+r^2-2rx)}\, \bA \cB^2,\quad
 \hat{A}^2_{21}=-\frac{ r^2 (-1+r x)}{1+r^2-2rx}\, \bB \cA \cB.\quad
\end{align}
%%%%%%%%%%%%%%%%%%%%%%%%%%%%%%%%%%%%%%%%%%%%%%%%%%%%%%%%%%%%%%%%%%%%%%%
The contributions coming from the symmetric bispectrum $\bispecAterm^{\rm(sym)}$, i.e., $A^n_{11}$, $A^n_{22}$, $\tilde{A}^n_{11}$, and $\tilde{A}^n_{22}$, coincide with those obtained in the auto-power spectrum case \cite{Taruya:2013my}, but others do not necessarily reproduce the previous results.  

\vspace*{0.5cm}

%%%%%%%%%%%%%%%%%%%%%%%%%%%%%%%%%%%%%%%%%%%%%%%%%%%%%%%%%%%%%%%%%%%%%%%
%%%%%%%%%%%%%%%%%%%%%%%%%%%%%%%%%%%%%%%%%%%%%%%%%%%%%%%%%%%%%%%%%%%%%%%
\subsubsection{$\Bterm$ term}
\label{subsec:B_term}
%%%%%%%%%%%%%%%%%%%%%%%%%%%%%%%%%%%%%%%%%%%%%%%%%%%%%%%%%%%%%%%%%%%%%%%
%%%%%%%%%%%%%%%%%%%%%%%%%%%%%%%%%%%%%%%%%%%%%%%%%%%%%%%%%%%%%%%%%%%%%%%

We first rewrite Eq.~(\ref{eq:B_term}) with 
%%%%%%%%%%%%%%%%%%%%%%%%%%%%%%%%%%%%%%%%%%%%%%%%%%%
\begin{align}
 \Bterm(k,\mu) &= \frac{(k\mu\,f)^2}{2} \cA\cB\,\int\frac{d^3\bfp}{(2\pi)^3} 
\Bigl[\tilde{F}_{\rm A}(\bfp)\tilde{F}_{\rm B}(\bfk-\bfp)+
\tilde{F}_{\rm A}(\bfk-\bfp)\tilde{F}_{\rm B}(\bfp)
\Bigr].
\label{eq:B_term2}
\end{align}
%%%%%%%%%%%%%%%%%%%%%%%%%%%%%%%%%%%%%%%%%%%%%%%%%%%
The integrand of this expression is symmetric under $\bfp \leftrightarrow \bfk-\bfp$. Then, as similarly done in the auto-power spectrum case \citep{TNS}, we expand the $\Bterm$ term in powers of $f$ and $\mu$, :
%%%%%%%%%%%%%%%%%%%%%%%%%%%%%%%%%%%%%%%%%%%%%%%%%%%
\begin{align}
 \Bterm(k,\mu)=\frac{k^3}{(2\pi)^2}
\sum_{n=1}^4\sum_{a,b=1}^2\,\mu^{2n}\,(-f)^{a+b} \int_0^\infty dr\int_{-1}^{1} dx\,\tilde{B}^n_{ab}(r,x)
\frac{P_{a2}(k\sqrt{1+r^2-2rx})\,P_{b2}(kr)}{(1+r^2-2rx)^a}. 
\end{align}
%%%%%%%%%%%%%%%%%%%%%%%%%%%%%%%%%%%%%%%%%%%%%%%%%%%
Note again that $r\equiv p/k$ and $x=(\bfp\cdot\bfk)/(pk)$. With the symmetric form of Eq.~(\ref{eq:B_term2}), the integral over $r$ and $x$ can be replaced with
%%%%%%%%%%%%%%%%%%%%%%%%%%%%%%%%%%%%%%%%%%%%%%%%%%%
\begin{align}
 \int_0^\infty dr\int_{-1}^{1} dx\,\,\longrightarrow\,\,
2 \int_0^\infty dr \int_{-1}^{{\rm Min}[1,\,1/(2r)]} dx.
\label{eq:range_integral}
\end{align}
%%%%%%%%%%%%%%%%%%%%%%%%%%%%%%%%%%%%%%%%%%%%%%%%%%%
This would help to improve the convergence of numerical integration, avoiding poles. The coefficient $\tilde{B}^n_{ab}$ is derived based on Appendix B of \cite{TNS}, and the results are summarized below:
%%%%%%%%%%%%%%%%%%%%%%%%%%%%%%%%%%%%%%%%%%%%%%%%%%%
\begin{align}
& \tilde{B}^1_{11}=\frac{r^2}{2}(x^2-1)\,\bA\bB\cA\cB, \quad
\tilde{B}^1_{12}=\frac{3r^2}{16}(x^2-1)^2\,\,\cA\cB(\bA\cB+\bB\cA), \quad
\tilde{B}^1_{21}=\frac{3r^4}{16}(x^2-1)^2\,\,\cA\cB(\bA\cB+\bB\cA), 
\nonumber
\\
&\tilde{B}^1_{22}=\frac{5r^4}{16}(x^2-1)^3\,\,\cA^2\cB^2,\quad
\tilde{B}^2_{11}=\frac{r}{2}(r+2x-3rx^2)\cA\cB\bA\bB,\quad
\tilde{B}^2_{12}=\frac{3r}{8}(x^2-1)(r+2x-5rx^2)\cA\cB(\bA\cB+\bB\cA),
\nonumber
\\
&\tilde{B}^2_{21}=\frac{3r^2}{8}(x^2-1)(-2+r^2+6rx-5r^2x^2)\cA\cB(\bA\cB+\bB\cA),\quad
\tilde{B}^2_{22}=\frac{3r^2}{16}(x^2-1)^2(-6+5r^2+30rx-35r^2x^2)\cA^2\cB^2,
\nonumber
\\
&\tilde{B}^3_{11}=0,\quad
\tilde{B}^3_{12}=\frac{r}{16}(4x(3-5x^2)+r(3-30x^2+35x^4))\cA\cB(\bA\cB+\bB\cA)\quad,
\nonumber
\\
&\tilde{B}^3_{21}=\frac{r}{16}(-8x+r(-12+36x^2+12rx(3-5x^2)+r^2(3-30x^2+35x^4)))\cA\cB(\bA\cB+\bB\cA),\quad
\nonumber
\\
&\tilde{B}^3_{22}=\frac{3r}{16}(x^2-1)(-8x+r(-12+60x^2+20rx(3-7x^2)+5r^2(1-14x^2+21x^4)))\cA^2\cB^2,\quad
\nonumber
\\
&\tilde{B}^4_{22}=\frac{r}{16}(8x(-3+5x^2)-6r(3-30x^2+35x^4)+6r^2x(15-70x^2+63x^4)+r^3(5-21 x^2 (5-15 x^2+11 x^4)))\cA^2\cB^2.\quad
\end{align}
%%%%%%%%%%%%%%%%%%%%%%%%%%%%%%%%%%%%%%%%%%%%%%%%%%%
Setting $\bA$, $\bB$, $\cA$ and $\cB$ to unity, 
the above expressions exactly coincide with those presented in Ref.~\cite{TNS}. 
%%%%%%%%%%%%%%%%%%%%%%%%%%%%%%%%%%%%%%%%%%%%%%%%%%

\onecolumn
\section{The relation between auto and cross power spectrum templates}
\label{sec:auto-x}
To show the relation and difference between the auto- and cross power spectrum templates in an explicit way, here we rewrite Eq.(\ref{eq:Pgkmu}) by introducing two sets of bias parameters 
\begin{equation}
\label{eq:b1}  \bI=\frac{\bIA+\bIB}{2}; \ \Delta \bI=\frac{\bIB-\bIA}{2},
\end{equation}
%%%%%%%%%%%%%%%%%%%%%%%%%%%%%%%%%%%%%%%%%%%%%%%%%%%
and
%%%%%%%%%%%%%%%%%%%%%%%%%%%%%%%%%%%%%%%%%%%%%%%%%%%
\begin{equation}
\bII=\frac{\bIIA+\bIIB}{2}; \ \Delta \bII=\frac{\bIIB-\bIIA}{2}.
\end{equation}
%%%%%%%%%%%%%%%%%%%%%%%%%%%%%%%%%%%%%%%%%%%%%%%%%%%
With Eq. (\ref{eq:bs2b3nl}), $b_{\rm s2}^{\A,\B}$and $b_{3{\rm nl}}^{\A,\B}$ can be written as
%%%%%%%%%%%%%%%%%%%%%%%%%%%%%%%%%%%%%%%%%%%%%%%%%%%
\begin{equation}
\bsA=-\frac{4}{7}\left(\bIA-1\right)=\bs+\frac{4}{7}\Delta \bI; \ \bsB=-\frac{4}{7}\left(\bIB-1\right)=\bs-\frac{4}{7}\Delta \bI,
\end{equation}
%%%%%%%%%%%%%%%%%%%%%%%%%%%%%%%%%%%%%%%%%%%%%%%%%%%
\begin{equation}
\bIIIA=\frac{32}{315}\left(\bIA-1\right)=\bIII-\frac{32}{315}\Delta \bI, \bIIIB=\frac{32}{315}\left(\bIB-1\right)=\bIII+\frac{32}{315}\Delta \bI.
\end{equation}
%%%%%%%%%%%%%%%%%%%%%%%%%%%%%%%%%%%%%%%%%%%%%%%%%%%
Substituting these new parameters into Eq. (\ref{eq:Pgdd}) gives
%%%%%%%%%%%%%%%%%%%%%%%%%%%%%%%%%%%%%%%%%%%%%%%%%%%
\begin{equation}
P_{{\rm g},\dd}^{\A\B}\left(k\right)=P_{{\rm g},\dd}\left(k\right)+\Delta P_{{\rm g},\dd}\left(k\right),
\end{equation}
%%%%%%%%%%%%%%%%%%%%%%%%%%%%%%%%%%%%%%%%%%%%%%%%%%%
where $P_{{\rm g},\dd}$ takes the form of the auto-power, \ie,
%%%%%%%%%%%%%%%%%%%%%%%%%%%%%%%%%%%%%%%%%%%%%%%%%%%
\begin{align}
P_{{\rm g},\dd}\left(k\right) & =\bI^{2}P_{\dd}\left(k\right)+2\bI\bII P_{{\rm b2},\updelta}\left(k\right)+2\bs\bI P_{{\rm bs2},\updelta}\left(k\right)\nonumber \\
& +2\bs \bII P_{\rm b2s2}\left(k\right)+2\bIII \bI \sigma_{3}^{2}\left(k\right)P_{\rm M}^{\rm L}\left(k\right)\nonumber \\
& +\bII^{2}P_{\rm b22}\left(k\right)+\bs^{2}P_{\rm bs22}\left(k\right)+N_{\A\B},
\end{align}
%%%%%%%%%%%%%%%%%%%%%%%%%%%%%%%%%%%%%%%%%%%%%%%%%%%
and
%%%%%%%%%%%%%%%%%%%%%%%%%%%%%%%%%%%%%%%%%%%%%%%%%%%
\begin{align}
\Delta P_{{\rm g},\dd}\left(k\right) & =-\left(\Delta \bI\right)^{2}P_{\dd}\left(k\right)-2\Delta \bI \Delta \bII P_{{\rm b2},\delta}\left(k\right)+\frac{8}{7}\left(\Delta \bI\right)^{2}P_{{\rm bs2},\updelta}\left(k\right)+\frac{8}{7}\Delta \bI\Delta \bII P_{\rm b2s2}\left(k\right)\nonumber \\
& -\frac{64}{315}\left(\Delta \bI\right)^{2}\sigma_{3}^{2}\left(k\right)P_{\rm M}^{\rm L}\left(k\right)-\left(\Delta  \bII \right)^{2}P_{\rm b22}\left(k\right)-\frac{16}{49}\left(\Delta  \bI \right)^{2}P_{\rm bs22}\left(k\right).
\end{align}
%%%%%%%%%%%%%%%%%%%%%%%%%%%%%%%%%%%%%%%%%%%%%%%%%%%
For $P_{{\rm g},\dt}^{\A\B}$, we find that both $\Delta  \bI $and
$\Delta  \bII $ vanish, so
%%%%%%%%%%%%%%%%%%%%%%%%%%%%%%%%%%%%%%%%%%%%%%%%%%%
\begin{align}
P_{{\rm g},\dt}^{\A\B}\left(k\right) = P_{{\rm g},\dt}\left(k\right).
\end{align}
%%%%%%%%%%%%%%%%%%%%%%%%%%%%%%%%%%%%%%%%%%%%%%%%%%%
To see how the A and B terms change under transformation of bias parameters,
we first rewrite them in the following form,
%%%%%%%%%%%%%%%%%%%%%%%%%%%%%%%%%%%%%%%%%%%%%%%%%%%
\begin{align}
A^{\A\B}\left(k,\mu\right) & =\mu^{2}f\left[A_{11a}\left(k\right) \bA   \bB   \cA +A_{11b}\left(k\right) \bA   \bB   \cB  \right]+\mu^{2}f^{2}\left[A_{12a}\left(k\right) \bB   \cA ^{2}+A_{12b}\left(k\right) \bA   \cB  ^{2}\right]\nonumber \\
& +\mu^{4}f^{2}\left[A_{22a}\left(k\right) \bB   \cA ^{2}+A_{22b}\left(k\right) \bA   \cA  \cB  +A_{22c}\left(k\right) \bB   \cA  \cB  +A_{22d}\left(k\right) \bA   \cB  ^{2}\right]\nonumber \\
& +\mu^{4}f^{3}\left[A_{23a}\left(k\right) \cA ^{2} \cB  +A_{23b}\left(k\right) \cA  \cB  ^{2}\right]\nonumber \\
& +\mu^{6}f^{3}\left[A_{33a}\left(k\right) \cA ^{2} \cB  +A_{33b}\left(k\right) \cA  \cB  ^{2}\right],
\end{align}
%%%%%%%%%%%%%%%%%%%%%%%%%%%%%%%%%%%%%%%%%%%%%%%%%%%
\begin{align}
B^{\A\B}\left(k,\mu\right) & =\mu^{2}\left[f^{2}B_{12}\left(k\right) \bA   \bB   \cA  \cB  +f^{3}B_{13}\left(k\right) \cA  \cB  \left( \bA   \cB  + \bB   \cA \right)+f^{4}B_{14}\left(k\right) \cA ^{2} \cB  ^{2}\right]\nonumber \\
& +\mu^{4}\left[f^{2}B_{22}\left(k\right) \bA   \bB   \cA  \cB  +f^{3}B_{23}\left(k\right) \cA  \cB  \left( \bA   \cB  + \bB   \cA \right)+f^{4}B_{24}\left(k\right) \cA ^{2} \cB  ^{2}\right]\nonumber \\
& +\mu^{6}\left[f^{3}B_{33}\left(k\right) \cA  \cB  \left( \bA   \cB  + \bB   \cA \right)+f^{4}B_{34}\left(k\right) \cA ^{2} \cB  ^{2}\right]\nonumber \\
& +\mu^{8}f^{4}B_{44}\left(k\right) \cA ^{2} \cB  ^{2}.
\end{align}
%%%%%%%%%%%%%%%%%%%%%%%%%%%%%%%%%%%%%%%%%%%%%%%%%%%
%%%%%%%%%%%%%%%%%%%%%%%%%%%%%%%%%%%%%%%%%%%%%%%%%%%
Setting $ \cA = \cB  =1$ as assumed in this paper, and eliminating
\textbf{$ \bA  $}, $ \bB  $ using Eq.(\ref{eq:b1}), we obtain,
%%%%%%%%%%%%%%%%%%%%%%%%%%%%%%%%%%%%%%%%%%%%%%%%%%%
\begin{equation}
A^{\A\B}\left(k,\mu\right)=A\left(k,\mu\right)+\Delta A\left(k,\mu\right),
\end{equation} with
%%%%%%%%%%%%%%%%%%%%%%%%%%%%%%%%%%%%%%%%%%%%%%%%%%%
\begin{align}
A\left(k,\mu\right) & =f\mu^{2}\left[A_{11a}\left(k\right)+A_{11b}\left(k\right)\right] \bI ^{2}\nonumber \\
& +f^{2}\left[\mu^{2}A_{12a}\left(k\right)+\mu^{2}A_{12b}\left(k\right)+\mu^{4}A_{22a}\left(k\right)+\mu^{4}A_{22b}\left(k\right)+\mu^{4}A_{22c}\left(k\right)+\mu^{4}A_{22d}\left(k\right)\right] \bI \nonumber \\
& +f^{3}\left[\mu^{4}A_{23a}\left(k\right)+\mu^{4}A_{23b}\left(k\right)+\mu^{6}A_{33a}\left(k\right)+\mu^{6}A_{33b}\left(k\right)\right],
\end{align}
%%%%%%%%%%%%%%%%%%%%%%%%%%%%%%%%%%%%%%%%%%%%%%%%%%%
\begin{align}
\Delta A\left(k,\mu\right) & =-f\mu^{2}\left[A_{11a}\left(k\right)+A_{11b}\left(k\right)\right]\left(\Delta  \bI \right)^{2}\nonumber \\
& -f^{2}\left[\mu^{2}A_{12a}\left(k\right)+\mu^{2}A_{12b}\left(k\right)+\mu^{4}A_{22a}\left(k\right)+\mu^{4}A_{22b}\left(k\right)+\mu^{4}A_{22c}\left(k\right)+\mu^{4}A_{22d}\left(k\right)\right]\Delta  \bI ,
\end{align}
%%%%%%%%%%%%%%%%%%%%%%%%%%%%%%%%%%%%%%%%%%%%%%%%%%%
\textbf{
	\begin{equation}
	B^{\A\B}\left(k,\mu\right)=B\left(k,\mu\right)+\Delta B\left(k,\mu\right),
	\end{equation}
}
%%%%%%%%%%%%%%%%%%%%%%%%%%%%%%%%%%%%%%%%%%%%%%%%%%%
\begin{align}
B\left(k,\mu\right) & =f^{2}\left[\mu^{2}B_{12}\left(k\right)+\mu^{4}B_{22}\left(k\right)\right] \bI ^{2}+f^{3}\left[\mu^{2}B_{13}\left(k\right)+\mu^{4}B_{23}\left(k\right)+\mu^{6}B_{33}\left(k\right)\right]\times2 \bI \nonumber \\
& +f^{4}\left[\mu^{2}B_{14}\left(k\right)+\mu^{4}B_{24}\left(k\right)+\mu^{6}B_{34}\left(k\right)+f^{4}\mu^{8}B_{44}\left(k\right)\right],
\end{align}
%%%%%%%%%%%%%%%%%%%%%%%%%%%%%%%%%%%%%%%%%%%%%%%%%%%
\begin{equation}
\Delta B\left(k,\mu\right)=-f^{2}\left[\mu^{2}B_{12}\left(k\right)+\mu^{4}B_{22}\left(k\right)\right]\left(\Delta  \bI \right)^{2}.
\end{equation}
%%%%%%%%%%%%%%%%%%%%%%%%%%%%%%%%%%%%%%%%%%%%%%%%%%%
Finally, the relation between the auto- and cross power spectrum templates is,
%%%%%%%%%%%%%%%%%%%%%%%%%%%%%%%%%%%%%%%%%%%%%%%%%%%
\begin{equation}
P_{\rm g}^{\A\B}\left(k,\mu\right)=P_{\rm g}\left(k,\mu\right)+\Delta P_{\rm g}\left(k,\mu\right),
\end{equation}
%%%%%%%%%%%%%%%%%%%%%%%%%%%%%%%%%%%%%%%%%%%%%%%%%%%
where
%%%%%%%%%%%%%%%%%%%%%%%%%%%%%%%%%%%%%%%%%%%%%%%%%%%
\begin{equation}
P_{\rm g}\left(k,\mu\right)=D_{\rm FoG}\left(k,\mu\right)\left[P_{{\rm g},\dd}\left(k\right)+2f\mu^{2}P_{\rm g,\dt}\left(k\right)+f^{2}\mu^{4}P_{\vv}\left(k\right)+A\left(k,\mu\right)+B\left(k,\mu\right)\right]
\end{equation}
%%%%%%%%%%%%%%%%%%%%%%%%%%%%%%%%%%%%%%%%%%%%%%%%%%%
is the auto-power spectrum, and
%%%%%%%%%%%%%%%%%%%%%%%%%%%%%%%%%%%%%%%%%%%%%%%%%%%
\begin{equation}
\Delta P_{\rm g}\left(k,\mu\right)=D_{\rm FoG}\left(k,\mu\right)\left[\Delta P_{{\rm g},\dd}\left(k\right)+\Delta A\left(k,\mu\right)+\Delta B\left(k,\mu\right)\right]
\end{equation}
%%%%%%%%%%%%%%%%%%%%%%%%%%%%%%%%%%%%%%%%%%%%%%%%%%%
gives the difference. 

The above calculation explicitly shows that the template of the cross power cannot be represented using that for the auto-power by redefining a single set of bias parameters.